# Color deconfinement and subhadronic matter: phase states and the role of constituent quarks

I. I. Royzen, E. L. Feinberg, O. D. Chernavskaya

**Contents**

**Abstract.** Major aspects of the subhadronic state of nuclear matter populated with deconfined color particles are reviewed. At high and even at rather low nuclear collision energies, this is expected to be a short-term quark-gluon plasma (QGP), but, seemingly, not only this. Emphasis is put on the self-consistency requirement that must be imposed on any phenomenological description of the evolution of a hot and dense nuclear medium as it expands (cools down) to the point where the final scattering of secondary particles starts. The view is argued and analyzed that massive constituent quarks should then play a major role at a certain cooling stage. A hypothesis is discussed regarding the existence of an intermediate stage (a valon plasma), allowing a consistent explanation of data on the mid-rapidity yields of various kinds of hadrons and direct dileptons (e$^+$e -pairs) in high-energy heavy-ion collisions.

## 1. Introduction

The modern understanding of possible extreme (subhadronic) states of nuclear matter reviewed below implies a description of the states that are only formed at a sufficiently high energy density (high temperature $T$ and/or high net baryon density $n_B$ - $n_{\bar{B}}$). Attempts to realize the proper conditions artificially involve collisions of two heavy atomic nuclei of sufficiently high energy, whereas in nature, they could be set in under high gravitational compression. The relevant physical processes are described within the framework of the quantum-statistical approach for many-particle systems with the QCD (quantum chromodynamics) interaction between particles. In accordance with QCD, the quarks, carrying so-called 'color' charges (three, rather than two charges as in electrodynamics), are the sources of the corresponding (color) field, while the gluons, being the quanta of the field, provide color exchange between interacting quarks and hence also carry color (more precisely, all the possible compositions of 'color' and 'anticolor'). For the problems under consideration, the crucial role is played by the lightest quarks u and d from which the 'ordinary' hadrons are built, and by the heavier quark s because the masses of the, c, b, and t quarks are far over the temperatures and/or Fermi energies typical of the phase transitions, and thus the corrections stemming from them are exponentially small.

A common belief has been established that the QCD Lagrangian should result in color confinement, i.e., that the quarks are always arranged to form some colorless compositions — the only kind of physical objects that can propagate freely. These are the hadrons, and therefore the radius of confinement is about the hadron size. This fundamental property is usually related to a vacuum condensate of gluons, which is formed under the 'ordinary' conditions such that its vacuum energy is lower than that of the naive 'empty perturbation' vacuum (as well as of the 'interquark vacuum' within the hadrons). The same can be said about the atomic nuclei because of the rarity of nucleons there: their proper (excluded) volume does not exceed 1/3 of the total nucleus volume.

But as the hadronic density (and hence the energy density of the nuclear substance) becomes sufficiently high, the nucleons first form a close packing and then begin to crush each other. As a result, colored particles locked inside them before lose their orientation to the parent hadrons, which are no longer the separated 'nuts', and start to propagate freely over the entire volume of such a medium, which may reach macroscopic sizes. In this sense, one can say that deconfinement of color (and of color-marked quark) occurs.

Under ordinary conditions (i.e., in the hadronic phase), the QCD vacuum is filled not only with the gluonic condensate, however. In addition, this vacuum must contain a quark–antiquark (chiral) condensate, which is a certain coherent superposition of states with different helicities. This is linked to the spontaneous chiral symmetry violation and is inherent in any theory of strong interactions where the axial isotriplet of π-mesons plays the role (almost) of the Goldstone bosons. It is very important (especially for the lightest quarks) that quarks can acquire a quite significant additional mass from their interaction with this chiral condensate.

There can be little doubt that both condensates should be destroyed sooner or later as the energy density increases, which results in the corresponding modification in the spectrum of excitations (just as with many other well-known symmetries — examples are ferromagnetism, superconductivity, electroweak interactions, etc.).

Thus, to all appearances, as the energy density increases, the nuclear substance should undergo two phase transitions — color deconfinement (due to the destruction of the gluon condensate) and chiral symmetry restoration (due to the destruction of the quark–antiquark condensate). However, while the former transition can proceed, in principle, without crucial modification of the quarks themselves as subhadronic excitations over the QCD vacuum, the second should affect them much more fundamentally: namely, it brings the masses of the lightest quarks (u and d) down to practically zero and also makes masses of the others correspondingly lower, thus turning the constituent quarks into current quarks (this term emphasizes that just these quarks enter the QCD Lagrangian and, hence, the expressions for quark currents).

Our understanding of the physical content of processes that go on in between hadronic and quark–gluonic states of matter depends crucially on answering the basic question: whether two above condensates are destroyed (occur) at the same time? We now clarify this. In the hadronic phase, with the color screening length given by

$$L_{cs} \simeq R_0 \simeq 1 \text{ fm},$$

the color confinement is maintained (where $R_0^{-1}$ can obviously be considered an order parameter), the chiral symmetry is broken, and even the lightest quarks have masses; at

$$L_{cs} \ll R_0 \simeq 1 \text{ fm},$$

when the nuclear matter is very hot and/or strongly compressed, color can propagate freely (no order parameter exists: $R_0^{-1} = 0$), the chiral symmetry is restored, and the lightest quarks are (nearly) massless. Now, we reformulate the above question as follows: is the matter stable in between ($L_{cs} < R_0$, but not $\ll R_0$)? And if yes, then what are the relevant peculiarities?

The point of view currently prevails that both transitions always proceed under the same conditions, i.e., coincide in time, and that no intermediate region exists at all. In other words, that the subhadronic state of nuclear matter that sets in as a result of color deconfinement turns at once into the chirally symmetric state due to immediate formation of the QGP. Below, this point of view is discussed in more detail, but special attention is paid to the alternative attitude — that the color deconfinement may occur at significantly lower heating and/or compression than the chiral symmetry restoration. Of course, the latter scenario necessarily implies that there is some (intermediate) color-conducting but still chirally broken phase, in which a quark can have a dynamical mass in addition to the current one (which directly enters the QCD Lagrangian); this is especially significant for light quarks. This hypothesis, as well as the special meaning of the dynamical mass scale, had been pronounced qualitatively long ago [1–3]. Moreover, the color confinement has been proven to precede the chiral symmetry restoration in the temperature scale [4] if these two transitions do not coincide. We prefer this scenario of nuclear matter phase evolution, first, because it is free of a number of inconsistencies that inevitably occur in the scenario that ignores the intermediate phase and, second, because the existing quite nontrivial experimental data on production of different hadron species and $e^+e^-$-pairs in the course of relativistic heavy ion collisions are described within the corresponding framework at least as successfully as in the current scenario. The review is accented on the comparison of the two aforementioned scenarios.

## 2. Retrospect, heuristic considerations, and QCD

Two apparently unrelated phenomenological ideas were put forward in the middle of the 1960s. The first one marked the beginning of subhadronic physics. It showed a possible way to systematize and arrange into acceptable order the extensive and seemingly rather chaotic hardonic 'zoo' under the assumption [5, 6] that all hadrons are 'built' of a minimal composition (as allowed by quantum numbers) of underlying, more elementary, particles — the massive u, d, and s quarks and the corresponding antiquarks combined in the proper way. Later on, these massive quarks Q (which could be called the heavier 'prototypes' of the current QCD-quarks, whose masses are about 330 MeV lower) were referred to as 'constituent quarks' (later also as 'valons' [7]); they played the basic role in the so-called additive quark model (AQM, see, e.g., [8–10]), which turned out to be surprisingly fruitful for understanding and sometimes even for quantitatively describing the soft hadronic interaction in a number of experiments. Then, along with advent of QCD, the notion of these 'model-motivated' quarks and the AQM itself gradually went out of use, although the relevant gap in the description of soft processes remained unfilled until now.

The second idea resulted in less significant consequences. On the face of it, a rather paradoxical thought came to mind about a limitation to the possible heating of the matter. This guess was based on the following heuristic consideration. The partition function of a kinetically and chemically balanced mixture of *ideal* gases with a spectral density $\rho(m)$ ($m$ being the particle mass) can evidently be expressed as

$$Z(T, V) \sim \exp\left[\frac{VT}{2\pi^2} \int_0^\infty m^2 K_2\left(\frac{m}{T}\right) \rho(m) \, dm \right],$$

where $T$ in the temperature, $V$ is the volume, and $K_2$ is the proper Bessel function, $K_2(m/T) \sim (T/m)^{1/2} \times \exp(-m/T)$ at $m/T \gg 1$. The hypothesis has been expressed [11] that the real gas of strongly interacting hadrons (including resonances) is equivalent (dual) to such a mixture; it has also been noticed (ibid) that the experimentally measured hadronic mass spectrum (although, within a rather narrow range of low masses) could be interpolated quite well by the function

$\rho(m) \sim \exp(m/T_0)$, where $T_0 \simeq 140-150$ MeV. If one could extrapolate this dependence up to the unlimited masses, then one finds that $Z(T, V) \to \infty$ as $T \to T_0$, and therefore this hypothetical hadronic gas would exist at $T < T_0$ only.

Taken literally, neither idea was complete and consistent. The former one — because it does not extend beyond particle phenomenology and particle grouping into so-called multiplets in accordance with their quantum numbers; no dynamical question — what is the mechanism of interaction between quarks — was elucidated (at that time, nothing was known about gluons). The latter one — because hadrons are extended particles (their form factors were measured experimentally), and the simplest estimate (similar to what is exemplified by Eqn (7), see below) shows that at $T \simeq T_0 \simeq 140-150$ MeV, practically no vacant space (the space not occupied by particle bodies themselves) would remain within such an 'ideal gas'. Thus, the vital discrepancy with the original hypothesis regarding the ideal gas is quite evident.

Nevertheless, these two ideas gave a great push to the development of both theory and experiment. The first one needs no additional recommendation — it has become a precursor for the consistent theory of strong interactions, i.e., for the QCD. The second one was suggestive for the guess that a phase transition from the hadronic phase (when color charges are locked within hadrons) to the deconfined phase (when color charge propagation is bounded by the extension of the medium under consideration only) is to be expected as the equilibrium nuclear matter is heated up to $T \simeq T_0$. This phase transition results from the increase in particle population density due to multiple production of hadrons and hadronic resonances (predominantly of pions). As a result, at sufficiently high temperatures, the mean interparticle spacing approaches the hadronic size itself, helping to trigger the color deconfinement.

Thus, the aforementioned 'limiting' temperature $T_0$, mysterious in its physical sense, turns into a conventional phase transition temperature. Of course, this temperature is not universal: in media with a nonzero net baryonic charge ($n_B - n_{\bar{B}} \neq 0$, for example, ordinary matter), a similar phase transition should occur even under 'low-temperature' (gravitational) compression as the baryonic charge density becomes sufficiently high. Moreover, the expected phase evolution of nuclear matter in the course of 'cold compression' seems to be most transparent in theoretical modeling [3]. This is exemplified in Fig. 1.

The phase that is just next to the ordinary substance [1] has been well-studied — this is dense nuclear matter. It forms when 'cold' compression of the nuclear matter results in close packing of the nuclei, leaving no vacant internuclear space (the excluded volume equals the total one), which forces them to start 'dissolving in each other'. This is undoubtedly the case within neutron stars. In the course of further compression, the nucleon packing density (approximately three times as high as within the nuclei) is reached at which the nucleons themselves come in contact and then start to 'dissolve in each other'. As a result, the colored particles freely propagate over the entire volume,[2] which means color deconfinement.

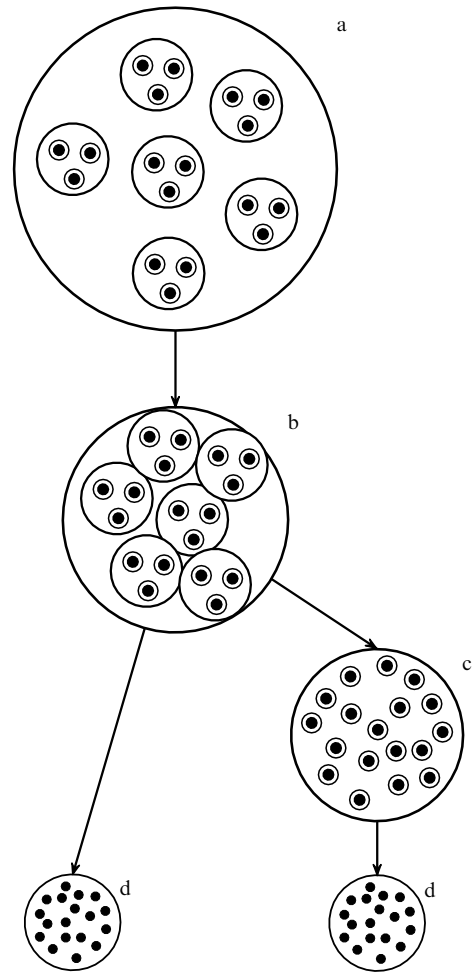

**Figure 1.** Schematic pattern of the conceivable ways for 'cold' nuclear matter to evolve along with gradual compression. The big circle refers to an atomic nucleus, the smaller ones to the hadrons, the smallest ones to the valons, and the dots to the current quarks. The left-hand branch refers to the coincidence of conditions for color deconfinement and chiral symmetry restoration (direct transition), while the right-hand branch implies that color deconfinement precedes chiral symmetry restoration (indirect transition, an intermediate phase is incorporated).

Two ways of evolution are conceivable. If the valons themselves are unstable, then color deconfinement (overlapping of nucleons) works as a trigger for their immediate turning into current quarks and thus results in formation of the chirally symmetric phase at once (direct transition b → d, the left-hand branch in Fig. 1). But if the valons Q(q) (here and below, q ≡ u, d) are rather stable quasiparticles with an effective radius $r_Q \ll r_N$ (where $r_N \simeq 0.8$ fm is the nucleon radius) — and AQM phenomenology implies this is reasonable and gives an estimate $r_Q \simeq 0.2-0.3$ fm [10] — then they can 'survive' as the compression and density grow quite significantly (in this connection, see estimate (7) below); this corresponds to the existence of intermediate phase 'c' (the right-hand branch in Fig. 1, which depicts the indirect transition b → c → d). Under further increase of compression (more intimate packing), the valons start to overlap and 'lose their coat'.[3] This process can develop either jump-like: until a certain time, the valons remain (almost) unchanged and then break down at once, turning into current QCD quarks and gluons, which should manifest itself as a sharp phase transition (probably even a first-order one) of matter into the chirally symmetric state; or they 'lose flesh' gradually and smoothly. In the latter case, a rather weak phase transition is most probably expected, being smeared in the phase plane to become a crossover.[4] Anyway, in both cases, one is dealing the intermediate phase suitable for the survival of 'liberated' valons.

---

[1] Nuclear compression is understood; the electronic structure of atoms plays no role here.
[2] Compared to the confinement scale, this volume is almost equally macroscopic for compression of a neutron star and of one or several heavy nuclei.

[3] The virtual gluons and quark–gluon pairs in the valon structure are meant.

Similar arguments seem to be quite reasonably applicable [2] to the treatment of head-on (central) collisions of heavy nuclei at relatively low energies, starting with the total CMS energy per nucleon pair (NN) $\sqrt{s_{NN}} \simeq 2.5 - 3$ GeV (the Lorentz factor $\simeq$ 1.5), when the temperature of the formed state is still rather low. Indeed, the nucleon mass (energy) density is only about three times as high as that of the nucleus, and therefore simple 'superimposing' (interpenetrating, occurrence in the same volume) of two identical nuclei, each of them relativistically contracted by about one and a half times, leads to the nucleonic mass (energy) density and the corresponding density of valon packing within the occupied volume, resulting in the valons freely propagating through the entire volume occupied by the overall nuclear substance. The conditions for this are only improved due to a noticeable number of pions that can be created in the course of nucleus 'stopping' (deceleration), which can result in an additional increase in the hadronic density and, evidently, favors color percolation even greater.

Certainly, this estimate is based on a rather abstract idealization, being thus purely illustrative. It is clear that this is the minimum estimate. The realistic color deconfining energy may occur significantly higher.[6]

Now, we turn from the above heuristic discussion to more exact treatment, although its conclusive and predictive power is also rather limited due to reasons to be explained shortly.

The QCD Lagrangian is written as

$$L(A,\psi) = -\frac{1}{4} F^a_{\mu\nu} F^a_{\mu\nu}$$
$$+ \sum_f \left[ i\bar{\psi}_f \gamma^\mu (\delta_\mu - ig\frac{\lambda^a}{2} A^a_\mu)\psi_f - m_f \bar{\psi}_f \psi_f \right], \quad (1)$$

where $A^a_\mu$ is the potential of the gauge gluon field that in QCD plays a role similar to that of photons in quantum electrodynamics (QED). A very important difference, however, is that the gluons are charged (colored) just like the quarks themselves. In this Lagrangian, $\psi_f$ is the field of a quark with the mass $m_f$ ($f$ is an additional quantum number of the quark - the so-called flavor, $f = 1, 2;..., N_f$ the quark color label being omitted), $g$ is the color charge responsible for the strength of interaction; $\lambda^a$ are twice the generators of the color group SU(3) in the fundamental representation (commonly known as the Gell-Mann matrices in this context), $a = 1, 2; ..., 8,$ and $f^{abc}$ are the corresponding structure constants (a totally antisymmetric tensor); see, e.g. [5,13].

'Coloration' of the gluon field manifests itself in the expression for its strength,

$$F^a_{\mu\nu} = \delta_\mu A^a_\nu - \delta_\nu A^a_\mu - gf^{abc} A^b_\mu A^c_\nu,$$

where the last term describes the gluonic self-action.

Nominally, the current masses of all six quarks enter Lagrangian (1) and, being nonzero, strictly speaking, manifestly violate its chiral symmetry. But the chiral properties of the Lagrangian actually depend almost completely on the lightest mass. The main point is that their current masses, $m_u \simeq 5$ MeV and $m_d \simeq 8$ MeV [13], which enter the QCD Lagrangian and are not caused by the color interaction,[7] are negligibly small compared to the typical QCD mass (energy) scale. As regards the (unstable) s quark, its current mass is $m_s \simeq 150$ MeV [13] and its weight factor equals to that of the u or d quark, and therefore its influence would result in corrections to all numerical estimates not exceeding $\simeq 0.5 \exp(-2m_s/T) \simeq 10\%$ even at the maximal expected temperature of the chiral phase transition, $T_c \leq 200$ MeV (see below), and steeply falls off as the temperature decreases. Thus, its appearance in the QCD Lagrangian cannot essentially affect the general pattern of phase transitions in the nuclear matter. This is all the more true for the heavier quarks. To this extent, Lagrangian (1) is considered to be approximately chirally symmetric, which is therefore the case for the corresponding so-called perturbative ('empty') vacuum.

Meanwhile, as was mentioned already, the phenomenological analysis shows [14 - 19] that the real physical QCD vacuum cannot be empty in the hadronic phase: it is populated with the gluon and chirally nonsymmetric quark condensates. The former provides color confinement, while the latter is linked to the (significant) breaking of the chiral invariance in this phase. Certainly, no perturbation theory enables us to describe both effectsÐthey are far beyond this approach. However, the necessity of breaking the chiral symmetry emerges unambiguously just from the fact that the (large) nucleon mass is not determined by the current masses of the quarks from which the nucleon is built and must remain practically unchanged even as their masses tend to zero. This is possible only if the π-meson is an (almost) Goldstone particle, whose mass $m_\pi$ is abnormally low (in the hadronic mass scale). This mass is known (see, e.g., [14]) to be found from the relation

$$f_\pi^2 m_\pi^2 = (m_u + m_d)\langle 0 | \bar{\psi}_L \psi_R + \bar{\psi}_R \psi_L | 0 \rangle, \quad (2)$$

where $f_\pi \simeq 90$ MeV is the coupling constant of the quark axial current to the π-meson and the last factor in the right-hand side is just the density of the chiral vacuum condensate [$\bar{\psi}_L(\bar{\psi}_R)$. and $\psi_L(\psi_R)$ are the respective creation and annihilation operators of quarks with left (right) helicity][8]. Certainly, a particle propagating through the vacuum of such a chiral structure can have no definite helicity, even if it undergoes no extra interactions. One can only conceive a *massive* particle in such a state because its helicity is well known not to be a relativistic invariant. It is therefore assumed that because of the presence of a helicity-violating

---

[4] As an analogy, one can take, for example, transition from an ordinary atomic substance to a fully ionized electromagnetic plasma.

[5] Of course, these qualitative considerations are insufficient for precluding the occurrence of a continuous crossover that actually smears the transition between the hadronic and chirally symmetric phases.

[6] The observations are probably indicative of qualitative changes in the features of interaction - they should undoubtedly accompany the color deconfinement - only at $\sqrt{s_{NN}} \geq 7$ GeV.

[7] This is also true for the current masses of all other quarks. For the u and d quarks, these masses are related to the nonzero π-meson mass.

[8] Thus, the vanishing of the π-meson mass is a signature of the chiral symmetry restoration.

quark–antiquark vacuum condensate, even the lightest u and d quarks behave as massive particles. The acquired mass

$$M = \langle 0|\bar{\psi}\hat{M}\psi|0\rangle = \langle 0|\bar{\psi}_L \hat{M}\psi_R + \bar{\psi}_R \hat{M}\psi_L|0\rangle \neq 0,$$

where $\hat{M}$ is the quark mass operator, is referred to as the dynamical mass. Unlike the current one, it is variable: at $T > 0$, it increases along with condensed particle density and always vanishes as soon as it does. [9] Sum rules for the reaction $e^+e^- \to$ hadrons in the vicinity of $\rho$- and $\omega$-resonances give the value 1.7 fm$^{-3}$ for the low-temperature physical vacuum density [19].

This all reasonably suggests that the constituent quarks (valons) Q in AQM can be identified with just these quarks with a dynamical mass. In the ordinary hadronic phase, this mass is then to be taken equal to about 1/3 of the nucleonic mass (for the light quark), i.e., $m_{Q(u)} \simeq m_{Q(d)} \simeq 330-340$ MeV and $m_{Q(s)} \simeq 480-500$ MeV.

The colored particles — either valons or current quarks, it does not matter — form the hadrons, and therefore attracting forces act between them. In addition, because no free quarks are observed, one can conclude that their separation requires infinite energy. [10] This fact is expressed by the words 'color confinement.' The reasoning is commonly adopted (but not verified explicitly — this point has yet to be proved in spite of numerous and lengthy endeavors [20]) that the quark color field suppresses the large-scale (nonperturbative) vacuum fluctuations of gluon and light quark fields that are responsible for the formation of the above-mentioned vacuum condensates and for making the vacuum energy lower than that of the empty perturbative vacuum. Because these are long-range fields (similar to the Coulomb one), it is possible that after the integration over the entire (infinite) volume, this suppression results in an infinite energy increase, which then shows that free color particles cannot exist. A similar role is played by the color field within the bounded volume of a hadron: it results in a finite increase in the interquark vacuum energy (see details in the bag model below). The energy density of the vacuum condensate is estimated [19] as

$$\varepsilon \simeq -\frac{9}{32}\frac{\alpha_s}{\pi}\langle 0|F^2|0\rangle \simeq -600 \text{ MeV fm}^{-3}, \quad (3)$$

where $\alpha_s = g^2/4\pi$. Here, the vacuum mean gluon field is estimated by comparing the theory and experimental data on $e^+e^-$-annihilation into $c\bar{c}$-states, the quark–antiquark condensate contribution being omitted because the relevant correction never exceeds 10%. The operator $\alpha_s F^2$ is renormalization-invariant [11] (has no anomalous dimension), and therefore the right-hand side of (3) is a fundamental (and essentially nonperturbative) constant in the theory.

Another very important peculiarity of Lagrangian (1) is the so-called asymptotic freedom. Unlike the above-mentioned properties, asymptotic freedom is of a purely perturbative nature and has been proved unambiguously [21, 22]. It shows that the value of $\alpha_s$ (as well as of $g$) is not a constant but decreases as the squared 4-momentum $q^2$ transferred in the course of the interaction increases: at large values of $q^2$,

$$\alpha_s(q^2) \simeq \frac{12\pi}{(33 - N_f)\ln(q^2/\Lambda^2)},$$

where $\Lambda \simeq 200$ MeV. Therefore, one can reasonably expect that along with further heating, quark–gluon matter in the QGP phase can become similar to an ideal gas: as the color screening length $L_{cs}$ becomes essentially lower than $R_0 \simeq 1$ fm, the long-range forces can convert into a mean color field with a rather low effective short-range residual interaction over its background, $V \sim r^{-1}\exp(-r/L_{cs})$, by far weaker than the particle mean kinetic energy. Lattice calculations (see below) seem to prove this expectation with rather good accuracy for the infinite medium with zero net baryon density. On the other hand, the well-known arguments [23] about applicability of the ideal-gas approximation at low temperatures (near and below the degeneracy temperature) to a fermion gas become invalid at high densities, because of the attractive interaction of particles and the possibility of bound state formation.

## 3. The phase plane: the current outlook

The above discussion is qualitatively summarized by the general scheme shown in Fig. 2, where the current understanding of possible phase transitions in nuclear matter is depicted on the $\mu_B T$-plane ($\mu_B$ is the baryonic chemical potential determined as if the nuclear medium had ideal gas properties). One can express almost no doubt that at least two phases exist — the hadronic phase H (bottom-left of the dashed curve 1 that bounds the color-confined domain) and the quark–gluon plasma (above the chiral transition curve 2). Below dashed curve 1, the substance consists of hadrons and nuclei, with quark–gluon degrees of freedom completely frozen. At very low temperatures, the gaseous and liquid phases are still distinguished (they are separated by the small-length curve *3* near the axis $T = 0$, in the vicinity of which atomic nuclei exist). Above curve *2*, on the contrary, no hadrons survive and the nuclear substance must be described in terms of quark and gluon degrees of freedom. The very existence of curves *1* and *2* is somewhat conventional: one or both phase boundaries may happen to be somewhat blurred (crossover), which does not then allow speaking of a phase transition in its standard meaning. This is seemingly just the case [24] with the chiral transition (curve *2*) in the vicinity of $\mu_B = 0$ (and therefore curve *2* is not extended to the vertical axis). Such behavior (crossover) may result from the non-vanishing (although very small) masses of the current u and d quarks, which, strictly speaking, prevents explicit current symmetry restoration even when no quark–antiquark condensate remains.

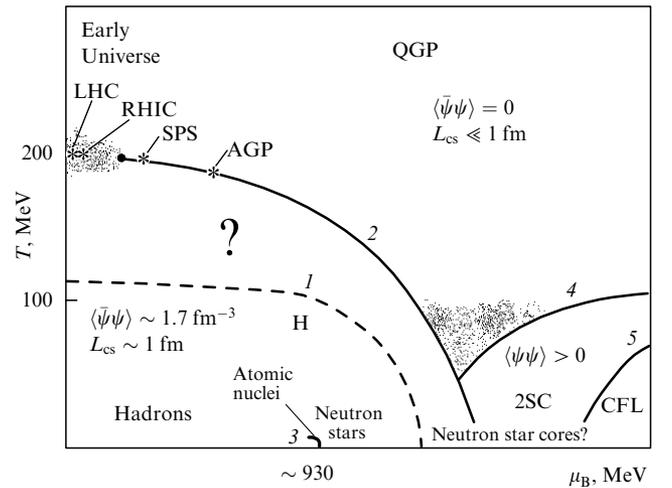

**Figure 2.** The phase plane $\mu_B T$. The curves 1, 2, 3, 4, and 5 refer to the respective phase transitions: the color deconfinement, the chiral transition, the liquid–gas transition in the hadron matter, the transition into color-superconducting diquark phases 2SC and CFL. Dot-shaded are the expected crossover domains. Star-marked are the expected chiral phase transition points at the averaged values of $\mu_B$ for the accelerators listed in the Table 1.

---

[9] The well-known nonrelativistic analogy of such behavior is ferromagnetism: the relation between the value of $\langle 0|\bar{\psi}\psi|0\rangle$ and the dynamical quark mass is the same as between the magnetization and external magnetic field; the temperature $T_c$ plays the role of the Curie temperature.

[10] Within the framework of potential (string) models, the interquark attraction strength is estimated to be about $1.2 \times 10^4$ kg m s$^{-2}$.

[11] More exactly, this is true for the operator $\beta(\alpha_s)F^2$, where $\beta(\alpha_s)$ is the Gell-Mann–Low function, which is of almost no significance for the point.

There is a certain reason (see, e.g., [25, 26]) to speak about formation of some specific subhadronic phase states at rather low temperatures (actually, $T \ll 100$ MeV is considered) but quite high baryon densities (to the right of curve *2*). These are the color-superconducting diquark phase states 2SC and CFL (below curves *4* and *5*, respectively), which set in due to the instability of the (ideal-gas) Fermi surface against the attraction between fermions (quarks), which is well known to result in Cooper pairing. Although no unquestionable theoretical (and, even more so, experimental) evidence of the formation at such states has yet been given, they seem to form even more evidently than in the similar case of the standard (electric) superconductivity, because, unlike electrons, quarks attract each other. The first state (2SC) is, most probably, a chirally symmetric color-superconducting state involving the pairing of u and d quarks in the isosinglet channel and the pairing of s quarks with each other;[12] the second state (CFL — color and flavor locked) seems to be a chirally broken state in which the color and flavor pairings of all the three, u, d, and s quark species are correlated nontrivially, such that a certain flavor pairing corresponds to a certain orientation in the color space. The symmetry properties of this state are expected to be exactly the same as for the hadronic matter compressed sufficiently high for making the nucleon and hyperon populations nearly equal. All mentioned features of ('cold') nuclear matter in this phase space domain may be relevant, evidently, for the processes within neutron stars. But all that is to be taken rather as reasonable guesses because color interaction is still quite powerful at the relevant densities (the value of $\alpha_s$ is not small because the typical momenta transferred are well below 1 GeV) and no reliable perturbative calculations can be performed.

In our opinion, the most controversial and interesting point (it has been noted above) is related to the presence or absence of the question-marked domain in between curves *1* and *2*, which is already unsuitable for hadronic survival (with the exception of π- and K-mesons, see below), but the chiral symmetry is still broken. In other words, the gluon vacuum condensate is partially destroyed there[13] by multiple pion production (high temperatures) and/or by compression (large values of $\mu_B$), while the quark–antiquark condensate and, therefore, massive valons are still in being.

---

[12] However, the question still remains, whether the color deconfinement and transition into this state occur at the same time, because the corresponding estimates [2, 26] of nuclear matter densities required for either one are close to each other by the order of magnitude only.

[13] Destroyed to the extent that the vacuum energy densities within and outside the hadrons become nearly equal.

Over the last 10–15 years, the theoretical and experimental investigations of extreme (subhadronic) states and the corresponding phase transitions in nuclear matter have become very extensive and multifarious, and their comprehensive review would be rather difficult. Another reason is that the number of questions is presently even greater than the number of reliable answers. Below, we confine ourselves to what is unambiguously related to the just formulated problem: what are the physical meaning of and peculiarities associated with the domain situated between curves *1* and *2*?

## 4. Theoretical models

### 4.1 Lattice calculations

The most reliable information about phase transitions in nuclear matter is commonly considered to emerge from direct lattice calculations.

The method is based on using the well-known mathematical technique of finding the quantum-statistical average in a system of particles that interact in accordance with QCD Lagrangian (1): as a result of the substitution $t \to i\tau$ $(0 \leqslant \tau \leqslant T^{-1})$, which is made in the functional integral representing the formal solution in ordinary space–time, this integral turns into the partition function

$$Z = \int DA_\mu D\psi D\bar{\psi} \, \exp\left[-\int_0^{1/T} d\tau \int_V d^3 x L(A_\mu, \psi, \bar{\psi})\right], \quad (4)$$

where integration over $\tau$ is performed with the periodic boundary conditions on the variable gauge field $A_\mu$ and with the antiperiodic ones on the Fermi fields $\psi$ and $\bar\psi$. Partition function (4) is calculated with the functional integration replaced by a multiple, but finite-dimensional integral (a lattice), which is then calculated by the Monte Carlo technique. In other words, the continuous space–time (with imaginary time) is replaced by a discrete space with the number of lattice sites $N_\sigma^3 \otimes N_\tau$ ($T = (N_\tau a)^{-1}$, $V = (N_\sigma a)^3$, $a$ is the lattice spacing, and $N_\sigma^3$ is the number of its space sites).

Theoretically, this might be considered not a mere model but a rather rigorous approach (and then no model considerations would be invoked at all), if one could ....

1. ... control the convergence to a certain limit as the lattice spacing tends to zero, or, at least, handle a sufficiently fine spacing (i.e., the lattice having a physically required number of lattice cells). Unfortunately, neither one nor the other is accessible for now — the latter simply because modern supercomputers[14] are still inadequate. The same can be said about the 'lattice' quark masses: one cannot maintain the correct relation between the masses of the current u and d quarks, on the one hand, and of the s quark, on the other, and therefore the calculations are performed for u and d quarks that are at least one order heavier than the real ones.

---

[14] A rough estimate is quite simple. The lattice spacing $a$ is required to be much smaller than a certain typical QCD scale which is reasonably equal to the inverse nucleon mass, i.e., $a \ll m_N^{-1} \simeq 0.2$ fm; at the same time, however, the lattice is supposed to cover distances at least about the diameter of a heavy nucleus, $\simeq 10$ fm. Therefore, a reliable quantitative result may be expected to be obtained only if the number of sites along each axis is well above $10^2$ and, hence, if the number of four-dimensional cells is quite larger than $10^8$. And this is the necessary condition only. The most optimistic estimate is that supercomputers that can handle integration over such sets will appear only in about 20 years.

2. ... perform lattice calculations that are reliable not only at $\mu_B = 0$ (although even in this case, allowance for fermions (quarks) results in great complications associated with computer shortcomings in anticommutative algebra [15]). An additional crucial problem at $\mu_B \neq 0$ is that no reliable method has been suggested (at least as of now) for making the results of the corresponding calculations gauge-invariant (i.e., sensible). True, in this connection, it is 'helpful' that the value of $\mu_B$ within the nuclear medium formed in an accelerator experiment is actually decreasing as the energy of nuclear collisions is increasing. However, at lower energies the problem remains.

Modulo all these shortcomings, the modern lattice calculations (at $\mu_B = 0$ and, possibly, near this point) show [24] that color deconfinement indeed occurs under sufficient heating of the nuclear medium, that the chiral symmetry restoration goes on at approximately the same temperature (within some 20% accuracy), and that this transition is, seemingly, a crossover [24] (see also Fig. 2). The main results of these calculations are shown in Figs 3–6. It is worth emphasizing that Figs 4–6 are quoted from paper [24], where no distinction between the color deconfinement temperature $T_d$ [16] and the chiral symmetry restoration temperature $T_c$ was made, and hence the intermediate phase was absent. That is why the deconfinement temperature was not mentioned at all, although just this temperature is actually the coordinate along the vertical axis. Equating these two temperatures, which are different in their physical sense, is based on a visual estimate and rather bold extrapolation of the relevant dependences (trends) obtained by numerical simulation (Fig. 3 [27] and Fig. 4 [24]).

The temperature dependence of the quark–antiquark condensate density $\langle 0|\bar{\psi}\psi|0\rangle$ and of the Wilson loop magnitude $L \propto \exp[-f_q(T)/T]$, where $f_q(T)$ is the free energy of an isolated quark, as well as of the relevant susceptibilities $\chi_L = \langle L^2 \rangle - \langle L \rangle^2$ and $\chi_{\bar{\psi}\psi} = \langle\langle 0|\bar{\psi}\psi|0\rangle^2\rangle - \langle\langle 0|\bar{\psi}\psi|0\rangle\rangle^2$ are shown in Fig. 3. Because of some shortcomings in the numerical simulation technique, one had to fix the ratio of the quark current mass to the temperature rather than this mass itself. Nevertheless, the simultaneous sharp decrease in both the chiral condensate density and the quark free energy followed by the thus much pronounced peak in the relevant susceptibilities (in the fluctuations of these thermodynamic functions) is usually considered to be conclusive for taking color deconfinement and chiral symmetry restoration as practically simultaneous phase transitions. It is not quite clear, however, how this statement jibes with the prediction of the same calculations that a crossover, but not the conventional phase transition, occurs at $\mu_B = 0$.

Figure 4 presents a collection of lattice results for the color deconfinement temperature $T_d$ obtained with various ways of taking the fermions into account (of choosing the fermion action). The ratio $T_d/m_V$ is shown as a function of the ratio $m_{PS}/m_V$, where $m_{PS}$ and $m_V$ are masses of the lightest pseudoscalar and vector mesons calculated within the same framework and are varied by assigning different values to the lightest quark current masses entering Lagrangian (1). The vertical straight line in Fig. 4a corresponds to the real mass ratio $m_{PS}/m_V = m_\pi/m_\rho$; the bold point on the ordinate axis ($T = T_c$) is to be taken as the most probable extrapolation (suggested by smoothing the $m_{PS}$-dependence of $T_d$) to the point $m_{PS} = 0$, which is correlated with the total disappearance of the helicity-nonconserving quark–antiquark condensate [see Eqn (2)] and thus to reestablishing the chiral symmetry. Unfortunately, the most realistic computation (the only point in Fig. 4b) is made (and can be made) at present only for the u and d quarks that are about one order heavier that the real ones ($m_{u,d} = 0.4T$, $m_s = T$).

The relevant values of $T_d$ obtained under these assumptions (and hence, by the postulated extrapolation, the values of $T_c$) are equal to $(173 \pm 8)$ MeV for $N_f = 2$ and $(154 \pm 8)$ MeV for $N_f = 3$. It is worth pointing out that the allowance for the third quark always makes this temperature lower by about 20 MeV. Furthermore, Figs 4a and b do not unambiguously give the dependence of $T_d$ on quark masses because the mass $m_V$ is by no means a scale independent of $m_{u,d}$. The more detailed analysis [24] showed that the 'lattice' temperature $T_d$ decreased quite slowly and almost linearly along with $m_{PS}$ (and thus with u- and d-quark mass) decrease. The trend found suggests that the correct value of this temperature can be still $(20-25)\%$ lower.

Figure 5 depicts the temperature dependence of pressure typical of the (first-order) phase transition that was computed for the pure gluodynamics (no quarks allowed at all), as well as when two and three quark flavors of the same mass or two light and one heavier quark flavors ($\mu_B = 0$) are allowed for, but with unrealistic current quark masses and the relation between them, $m_q = 0.4T$, $m_s = T$ (see also the caption to Fig. 4). The physical content of this transition is dequenching color degrees of freedom. But the transition temperature is noticeably smeared, which is indicative of a crossover. In this figure, $p_{SB}$ denotes the Stephan–Boltzmann pressure of the relevant ideal gas,

$$\frac{\varepsilon_{SB}}{T^4} = \frac{3p_{SB}}{T^4} = \frac{\pi^2}{30}\left(2 \times 8 + 2 \times 3 \times 2 \times \frac{7}{8} \times N_f\right),$$

where the numerical factors in the right-hand side take account for 8 gluons with two possible polarizations plus quarks and antiquarks of 3 colors with two possible polarizations, the factor 7/8 reflects the distinction between Fermi and Bose statistics, and $N_f$ is the number of flavors under consideration. One can see that the gas is not ideal: its pressure remains about 20% below $p_{SB}$ at least up to quite high temperatures $T \gg T_c$.[17] (We recall that the authors of [24] equated the temperatures $T_d$ and $T_c$.)

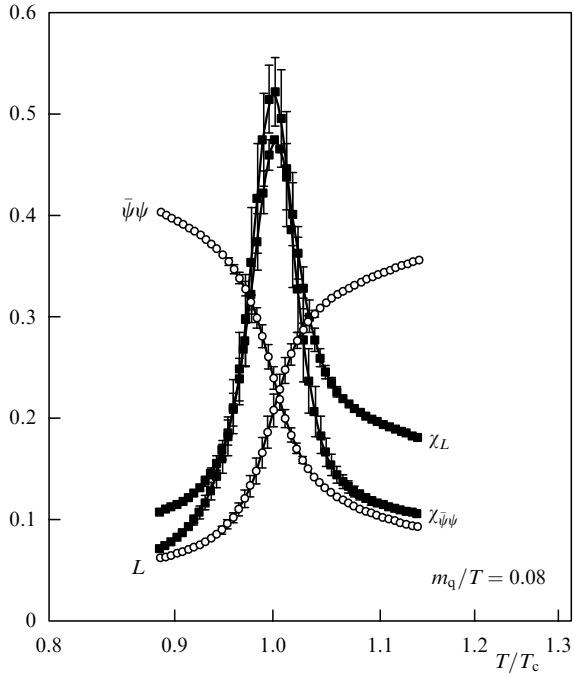

**Figure 3.** The lattice results (at $\mu_B = 0$) for temperature dependences of the chiral condensate density $\langle 0|\bar{\psi}\psi|0\rangle$ and Wilson loop magnitude $L \propto \exp[-f_q(T)/T]$ ($f_q$ is the free quark energy), as well as of the relevant susceptibilities $\chi_{\bar{\psi}\psi} = \langle\langle 0|\bar{\psi}\psi|0\rangle^2\rangle - \langle\langle 0|\bar{\psi}\psi|0\rangle\rangle^2$ and $\chi_L = \langle L^2 \rangle - \langle L \rangle^2$ near the chiral transition and deconfinement temperatures, respectively. This is considered to be indicative of the coincidence of these temperatures.

---

[15] One has to do this part of the work 'by hand' and then the computer is instructed to calculate the determinants of the well-known high-rank matrices. Remarkable progress has been achieved in this skill over recent years.

[16] The same symbol $T_d$ is used below for the hadronization temperature, because the deconfinement and hadronization temperatures have the same physical meaning. The latter one is used, as a rule, when speaking about cooling a medium down, and that is why it is even somewhat more relevant in the context of evolution in the course of heavy nucleus interactions.

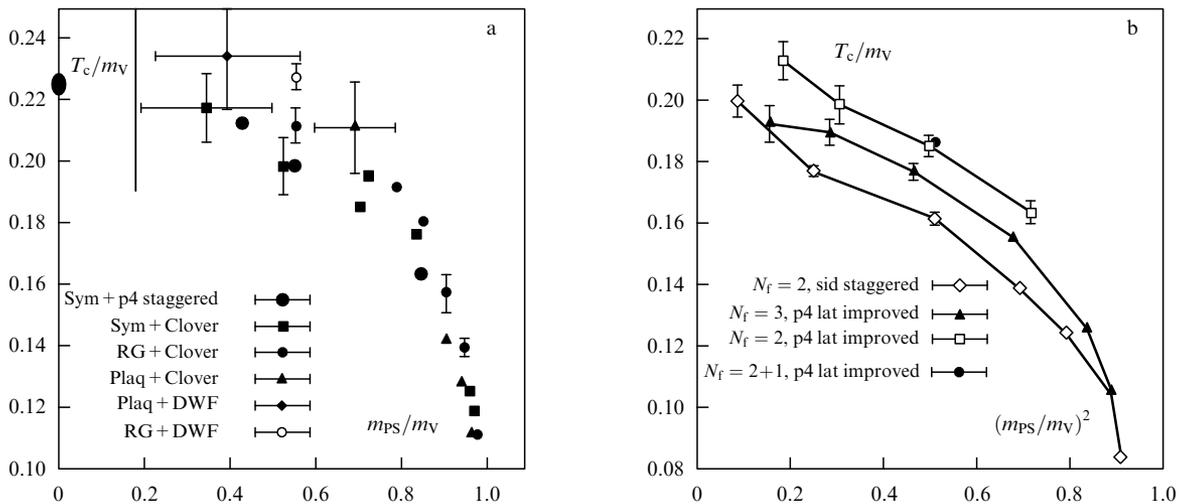

**Figure 4.** The lattice results for deconfinement (hadronization) temperature as a function of the ratio $m_{PS}/m_V$: (a) — with the fermion action included using different methods (see [24] and references therein), only two flavors, u and d quarks, are taken into account ($m_s = \infty$); (b) — comparing the results of different calculation versions for two flavors ($m_s = \infty$) to the calculation for three quark flavors of the same mass or two light flavors, u and d quarks, and one heavier flavor, $m_{u,d} = 0.4T$, $m_s = T$, ($N_f = 2 + 1$).

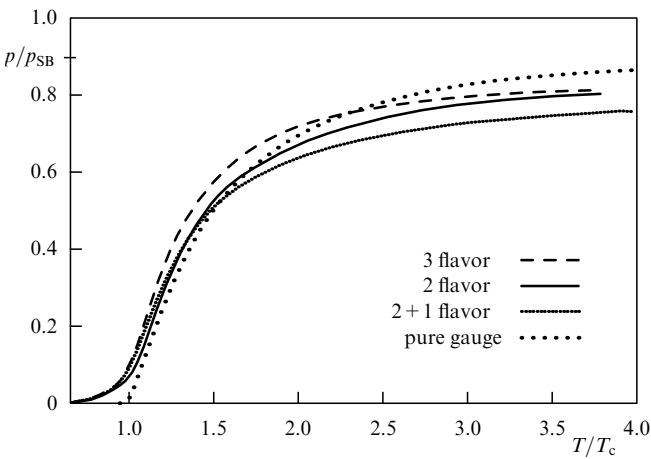

**Figure 5.** Lattice results for temperature dependence of pressure for pure gluodynamics, as well as for two and three flavors of the same mass and for two light and one heavier flavor ($\mu_B = 0$), albeit with unrealistic current quark masses and the relation between them: $m_{u,d} = 0.4T$, $m_s = T$; $p_{SB}$ is the Stephan–Boltzmann pressure of the corresponding ideal gas.

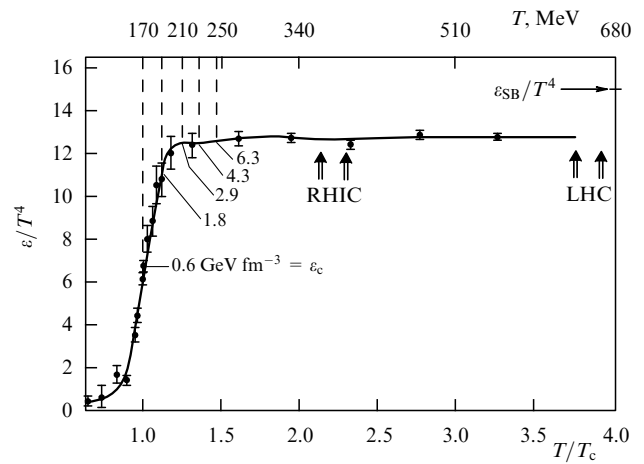

**Figure 6.** Lattice results for temperature dependence of the energy density with three 'light' quarks of the same mass ($\mu_B = 0$). One can see that color deconfinement occurs near the energy density $\simeq 0.6$ GeV fm$^{-3}$.

In Fig. 6, the temperature dependence of the energy density $\varepsilon$, typical of a phase transition (a first-order one, cf. Fig. 5), is shown for three 'light' flavors of the same mass at $\mu_B = 0$. This dependence indicates that color deconfinement is to be expected at $\varepsilon = \varepsilon_c \simeq (6 \pm 2)T_c^4 \to (0.6 \pm 0.2)$ GeV fm$^{-3}$ and $T_c \simeq 170$ MeV (the assumptions are the same as in Fig. 5). One can also see that at $T > T_c$, the predicted energy density is still noticeably lower than $\varepsilon_{SB}$, but the relation $\varepsilon/p \simeq \varepsilon_{SB}/p_{SB}$ already becomes valid at $T/T_c > 2-2.5$. The vertical arrows show the estimates of the initial temperature of the nuclear medium formed by collisions of heavy nuclei at the top energy of RHIC/BNL and LHC/CERN accelerators.

---

[17] This is seemingly associated with an interaction, typical of this medium, caused by some nonperturbative chromomagnetic fields (so-called 'magnetic confinement').

## 4.2 Bag model

Phenomenological models of the type of the MIT bag [28] are used to describe phase states and transitions in nuclear matter. Originally, this model was designed for describing just hadrons and their spectra, but it is evidently related to certain equations of state (EoS) of nuclear matter.

The MIT bag model is based on the idea that the fields forming the physical vacuum are forced out from the hadrons, which results in locking the quarks within the hadrons due to an extra pressure $B$ from outside, irrespective of whether this field is ejected totally or not;[18] this pressure is called the bag constant and is considered a free parameter of the model.

The intrinsic energy (mass) of a hadron is actually assumed equal to the sum of the corresponding 'extra vacuum energy' $(4\pi/3)Br_H^3$, where $r_H$ is the hadron radius, and the kinetic energy of massless quarks, $\sim n/r_H$, where $n = 2$ for mesons and $n = 3$ for baryons. The radii and masses of hadrons are then obtained by minimizing this sum with respect to $r_H$. The extra pressure $B$ and quark Fermi pressure are balanced to provide stability of the hadron in total.

The values of $B$ that allow a realistic description of the hadronic spectrum under some reasonable assumptions are in the range $B_{MIT} \approx 50-100$ MeV fm$^{-3}$, which is about one order lower than the pressure of the physical vacuum field that was estimated in a number of models [12, 18, 19, 29] to be $p^{vac} = -\varepsilon^{vac} \simeq 0.5-1$ GeV fm$^{-3}$ [this is also to be compared with estimate (3)]. This fact shows that the vacuum field is only partially (by about 10%) forced out from the hadrons and, therefore, that the interquark space within the hadrons cannot be considered the ideal perturbative vacuum, and that the nuclear medium in a hadron cannot be treated as a QGP.

This interpretation has given ground for the hypothesis that an extra stable intermediate phase, the so-called valonic or Q-phase,[19] can be formed in addition to the hadron matter and QGP, in which the physical vacuum is partially ejected from the entire volume occupied by the nuclear medium and thus the massive (constituent) quarks (valons) become free to propagate through this hadron-like vacuum. This has led to the idea of three-phase states of nuclear matter.

Three-phase nuclear matter is described by using the partition functions $Z_j(T, \mu_B, V)$ for each phase that depend on $T$, $\mu_B$, and the occupied volume $V$. Within the approximation of the ideal gas of point-like particles, these functions are given by

$$\ln Z_j^0(T, \mu_B, V)$$
$$= -\ln Z_j^{vac} + \frac{V}{T}\sum_i \frac{G_i^B}{6\pi^2} \int \frac{k^4 \, dk}{\sqrt{k^2 + m_i^2}}$$
$$\times \frac{1}{\exp\left(\frac{\sqrt{k^2+m_i^2}}{T}\right) - 1} + \frac{V}{T}\sum_i \frac{G_i^F}{6\pi^2} \int \frac{k^4 \, dk}{\sqrt{k^2 + m_i^2}}$$
$$\times \left[\frac{1}{\exp\left(\frac{\sqrt{k^2+m_i^2}-\mu_{Bi}}{T}\right) + 1} + \frac{1}{\exp\left(\frac{\sqrt{k^2+m_i^2}+\mu_{Bi}}{T}\right) + 1}\right], \quad (5)$$

where $G_i^{B,F}$, $m_i$, and $\mu_{Bi}$ are the known combinatorial factors, masses, baryonic chemical potentials, and partition functions, respectively, for Bose and Fermi species of the type $i$ in the $j$th phase ($j$ denotes the hadronic (H), valon (Q), and QGP phases). Summation goes over the species that dominate in the phase considered. It is worth emphasizing that a certain fraction of pions, which play the role here of Goldstone particles, is allowed for in the intermediate phase, besides the valons themselves. The term $\ln Z_j^{vac}$ in (5) is equal to $-(V/T)B_{QGP}$ in the QGP-phase or to $-(V/T)B_Q$ in the Q-phase and describes an effective interaction with the vacuum ensuring confinement of quarks or valons within the respective media.

One significant point has yet to be mentioned. The ideal gas approximation is rather reasonable for the QGP phase, whereas it is somewhat problematic for the Q phase and invalid for the H phase in the temperature interval of interest: some essential corrections must be made in the last case, bearing in mind the hadronic size $r_H \simeq 0.8$ fm and the sufficiently high hadron number density at the relevant temperatures They are usually introduced by using van der Waals type models with excluded volume (for details see, [3, 38]).

The EoSs $p_j(T, \mu_B, V) \equiv (T/V) \ln Z_j$ for each phase enable us to find the phase equilibrium points. The equilibrium condition $p_i = p_j$ of two arbitrary phases implies that three types of phase equilibrium curves can occur:

(1) the transition corresponding to *deconfinement of valons* as the temperature increases (*hadronization* in the course of cooling down), H $\leftrightarrow$ Q, at $T = T_d$, followed by (preceded) by

(2) the *chiral* transition Q $\leftrightarrow$ QGP at $T = T_c$;

(3) the *direct* transition H $\leftrightarrow$ QGP at $T = T_d = T_c$.

Of course, which one of the above patterns is predicted by the bag model relates intimately to the choice of the parameters $B_Q$ and $B_{QGP}$. At the same time, the physically admissible domain of their values is rather large. The large majority of works in which a possible role of the intermediate phase was analyzed has been coordinated with lattice calculations, which seemed quite indicative of the deconfinement and chiral transition coincidence for the baryon-neutral nuclear matter (i.e., at $\mu_B = 0$). This results in an additional constraint on the parameters: $B_Q/B_{QGP} \simeq 0.3$, and therefore only one of them remains free [30–33, 38].

But it seems more reasonable to choose the values of the parameters $B_j$ following their physical meaning, which gives $B_Q \equiv B_{MIT} \simeq 50-100$ MeV fm$^{-3}$ and $B_{QGP} \equiv p^{vac} \simeq 0.5-1$ GeV fm$^{-3}$. Actually, they were chosen [3] to be $B_{QGP} = 500$ and $B_Q = 50$ MeV fm$^{-3}$. This choice of the parameters resulted in the following estimates for the baryon-neutral nuclear matter: $T_d \simeq 140$ MeV and $T_c \simeq 200$ MeV.

The phase diagrams for the two parameter choices are shown in Fig. 7. One can see that the first version leaves practically no room for the intermediate phase, whereas the second clearly indicates quite a large temperature interval ($\Delta T \simeq 50$ MeV) where the Q-phase dominates. Certainly, the width of this interval depends on the model parameters, but it does not disappear even at $\mu_B = 0$.

We note that these particular models predict first-order phase transitions in nuclear matter characterized the appearance of the so-called mixed phase, which is a long-term fractal composition of the 'old' and 'new' phases at the same temperature, the 'old' one being gradually exhausted. This implies that in the course of evolution, more precisely, as the fireball cools down, the nuclear medium should spend a long time either as a mixture of hadrons and QGP bubbles (the first version) or as a mixture of hadrons and valonic bubbles (the second version). In the latter case, valons no doubt crucially affect the secondary hadron spectrum.

---
[18] If it were ejected totally, the inner hadronic vacuum would become perturbative, and therefore the u and d quarks within hadrons would be the current ones. This is just what was assumed originally, but it was soon realized that the hadronic sizes and spectra could not be described within such a framework.

[19] This is the notation adopted in the relevant works. We use it in this section only; in the rest of the paper, essentially the same state is referred to more precisely as the Q$\pi$K-phase.

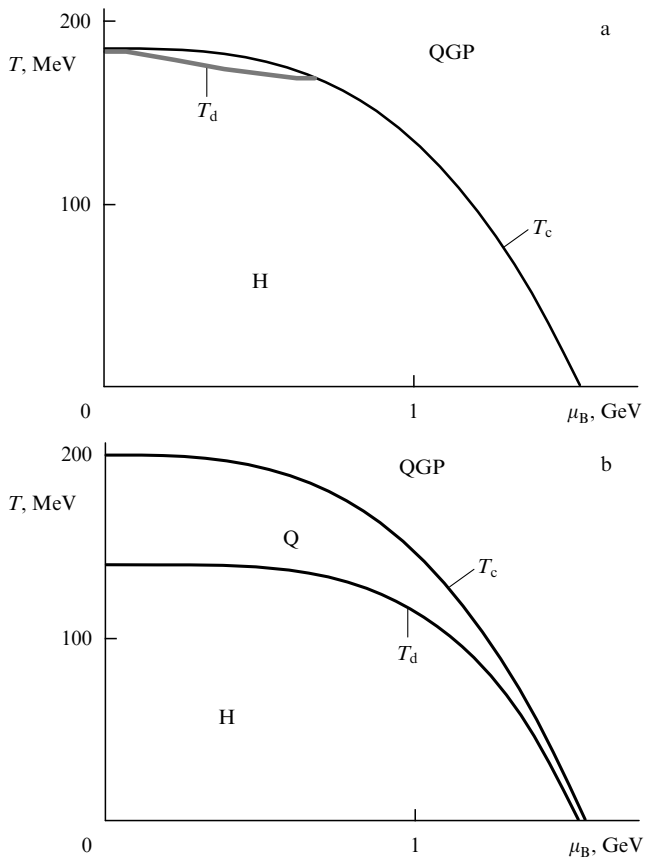

**Figure 7.** Phase plane for two versions of the bag model: (a) $B_{QGP} = 500$ and $B_Q = 170$ MeV fm$^{-3}$ (the parameters at which color deconfinement and chiral symmetry restoration occur simultaneously); (b) $B_{QGP} = 500$ и $B_Q = 50$ MeV fm$^{-3}$ (the parameters of the three-phase model based on physical arguments detailed in [3]).

### 4.3 Interim results: what do the theoretical models teach us?
Resuming the aforesaid in this section, one can conclude that

1. Both approaches predict the color deconfinement as well as the chiral phase transition, but the latter is more pronounced in the bag model. Bearing in mind all the above reservations, one can nevertheless agree that the available lattice calculations indicate the simultaneity of these changes in the nuclear matter state, whereas in the bag model, this question remains open: namely, an intermediate phase is clearly predicted to appear under a quite reasonable choice of parameters.

2. The bag model predicts the chiral phase transition temperature $T_c \simeq 200$ MeV (at $\mu_B = 0$, whereas the most realistic lattice calculations are only compatible with $T_c \simeq 150$ MeV (if one agrees that $T_c = T_d$!!) and a trend can be seen in which this temperature decreases as the 'lattice' matter approaches the real one.

3. In contrast to the lattice calculations, the bag model allows handling the media with chemical potentials $\mu_B \neq 0$.

Thus, one can see both a certain similarity and a quite noticeable quantitative and, maybe, even qualitative distinction in the predictions of these two models. Nevertheless, they are useful: having no better tools, one can appeal to them in making estimates and phenomenological speculations, respecting these models as a helpful, if not quite reliable, reference base.

## 5. Man-made subhadronic matter?

The models discussed in the preceding section are in fact static, i.e., they do not imply medium evolution caused by inner forces. But the most interesting problem is the theoretical understanding of the heavy nucleus collisions because they provide a unique (at least for now) chance to obtain color deconfinement and to have QGP at the laboratory. However, a number of questions linked to the nonstationary nature of the process are put forward in this connection, and thus the very statement that equilibrium subhadronic states, at least transient, must necessarily be formed is no longer undeniable. Everything that follows is to be taken modulo this problematic character. Primary attention should therefore be paid to searching for some indubitable signals that are directly linked only to such a state. We only touch on this problem casually in subsection 5.4.

### 5.1 A general view of the process
In a central collision of two heavy nuclei, the nucleons forming each of them have to overcome the formation of the nucleons they face. Some of them are fortunate enough to break through, being, maybe, pretty well 'wounded' (excited) but still standing against the catastrophic disintegration, i.e., preserving their colorlessness. Others (those that were crushed totally) produce randomly moving colored particles that cannot escape a bounded spatial domain unless and until they recombine to form certain stable blanched states — most predominantly, these are the hadrons and hadronic resonances (color confinement). This process takes a quite long time compared to the duration ($\simeq 1$ fm) of the 'prompt' interaction. Therefore, they decelerate, dissipating their energy to produce a large number of new colored particles and to heat (thermalize) the whole medium. As a result, three kinematic regions can be singled out along the whole pseudo(rapidity) interval allowed: the forward and backward wings of fragmentation and the mid-rapidity blob of nuclear matter, which is usually referred to as the fireball, see, e.g., Fig. 8 loaned from [27]. If identical nuclei collide, then this fireball is CMS-symmetric (but not isotropic) in all its parameters (in particular, its center of inertia is practically motionless). It follows from the experimental data that the higher the collision energy, the more numerous the nucleons that are 'fortunate' enough to break through (remaining colorless) to appear in the nucleus fragmentation region, taking the corresponding baryonic charge away with them. The fireball then becomes hotter and loses its baryon density. Specific signals indicating transient color deconfinement and creation of subhadronic states of nuclear matter must be sought in this range.

Various estimates show [34, 35] that the SPS/CERN-accelerator top energy ($\sqrt{s_{NN}} \simeq \sqrt{2mE_L} \simeq 17-19$ GeV/NN) is probably high enough for 'jumping' over curve 2 in Fig. 2, and, all the more, it is the case at energies at the RHIC/BNL accelerator put into operation in the middle of 2000 (see

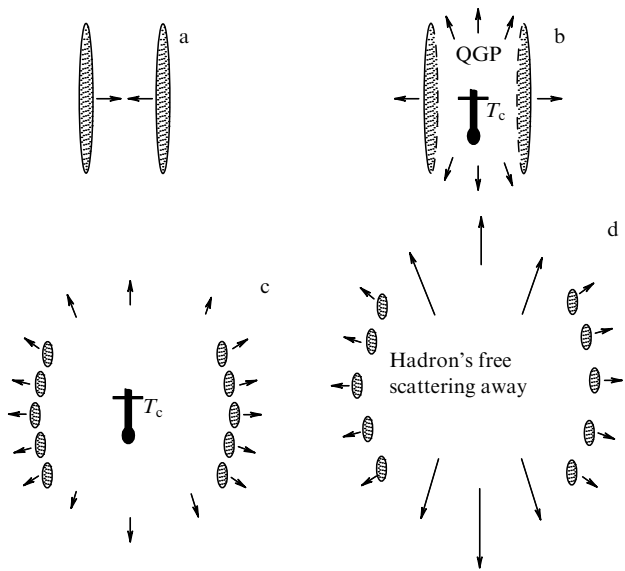

**Figure 8.** Schematic view on the successive stages of the central interaction of two identical heavy nuclei [27] (if an intermediate phase is allowed for, stage (c) splits into two, $T_c > T > T_d$ and $T_d > T > T_f$, the respective Q$\pi$K- and H-phases).

**Table 1.** Experimental tools for studying high-energy nuclear collisions.

| Accelerator | Put into opera-tion | Type | Beam | $\sqrt{s_{NN}}$, GeV | $\varepsilon_0^{AB}$, GeV fm$^{-3}$ |
|---|---|---|---|---|---|
| BNL–AGS | 1986 | Fixed target | $^{28}$Si | 5 | 0.7 |
| CERN–SPS | 1986 | Fixed target | $^{16}$O, $^{32}$S | 19 | 1.6 |
| BNL–AGS | 1992 | Fixed target | $^{197}$Au | 5 | 1.5 |
| CERN–SPS | 1994 | Fixed target | $^{208}$Pb | 17.5 | 3.7 |
| BNL–RHIC | 2000 | Collider | $^{197}$Au | 200 | 7.6 |
| CERN–LHC | ~2007 | Collider | $^{208}$Pb | 5500 | 13 |

*Note.* Heavy nuclei are used only in the fixed target ($A > 200$); the collider beams are symmetrical. $\varepsilon_0^{AB}$ is the calculated value of the initial energy density.

Table 1); more ambiguous is the answer to the question what happens at AGS/BNL, where $\sqrt{s_{NN}} \simeq \sqrt{2mE_L} \simeq 5$ GeV/NN?

Thus, there is ground to believe that short-term restoration of chiral symmetry and formation of QGP are quite probable in the relevant experiments, although there is still no reliable evidence that this is indeed the case.

The fireball then undergoes expansion and cooling, which leads to hadronization and, eventually, to the entire disappearance of the interaction. The temperature at which this occurs and the hadrons start scattering away freely, $T_f \simeq 100-110$ MeV, is usually referred to as the kinetic freeze-out temperature. It is important that this temperature is linked almost directly to the experimentally observed transverse momentum distributions of the final hadrons.

The general pattern of evolution is shown schematically in Fig. 9. The physical mechanism of this process is described by thermohydrodynamic models (see, e.g., [36, 38]). The basic assumption is that after thermalization of a nuclear medium (which takes only a short time from the beginning of the interaction), the hot substance starts evolving in accordance with the equations of relativistic hydrodynamics [37,38] for the energy–momentum tensor $T^{\mu\nu}(x)$; these equations express the local conservation laws (neglecting dissipation),

$$\partial_\mu T^{\mu\nu}(x) = 0,$$

$$T^{\mu\nu}(x) = [\varepsilon(x) + p(x)]\, u^\mu(x)\, u^\nu(x) + p(x)\, g^{\mu\nu}, \qquad (6)$$

where $x$ is the space–time coordinate, $\varepsilon(x)$ and $p(x)$ are the local energy density and pressure in the proper system of a given element of volume, respectively, $u^\mu(x)$ is its velocity 4-vector, and $g^{\mu\nu}$ is the metric tensor ($g^{00} = -1$, $g^{ij} = 1$). Of course, the solution of these equations depends on the initial conditions, which should be formulated making use of a certain physical reasoning. The Bjorken scenario [39], which is currently most popular, incorporates even some additional simplifying assumptions.

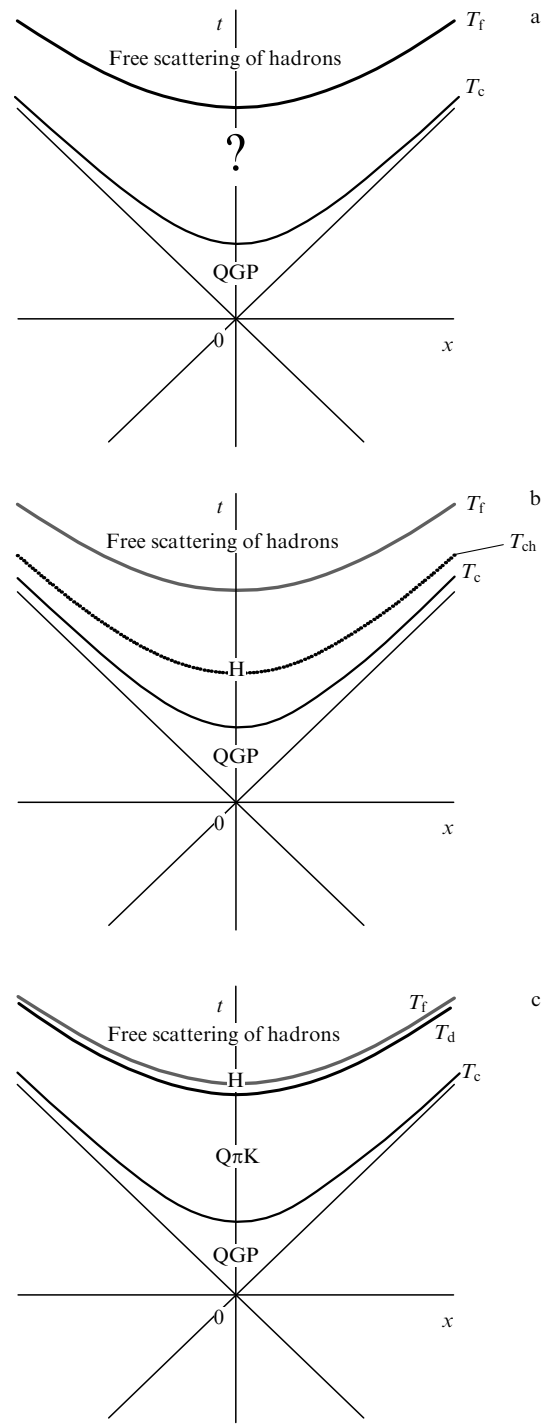

**Figure 9.** The space–time pattern of fireball evolution from the very beginning of interaction ($t = 0$) until the produced hadrons are scattered (the interaction between them terminates at $T = T_f$): (a) — the general appearance, (b) — the direct transition scenario, QGP → (H) (the curve $T = T_{ch}$ refers to the early chemical freeze-out of hadrons), (c) — the intermediate-phase scenario QGP → Q$\pi$K → (H), the curve $T = T_c$ refers only to the chiral transition between the phases QGP and Q$\pi$K. The hadronization and kinetic freeze-out temperatures $T_d$ and $T_f$ are quite close to each other.

Furthermore, Eqns (6) are not complete: to describe the evolution of a certain medium, the relevant *equations of state* must be incorporated that interrelate the thermodynamic parameters ($p, \varepsilon, T, \mu_B$). These equations can only be postulated.[20] If some discontinuous equations (for example, a jump of the parameter $B$ in the bag model) are used within this approach, then a pattern typical of evolution with phase transitions emerges. In this case, depending on the parameter choice, either the only (direct) transition QGP → H or two successive transitions QGP → Q$\pi$K → H occur. Correspondingly, the time-sweeps of evolution differ from each other.

Because of the obvious lack of predictive power of the theory — the lattice calculation and bag models are meant — the choice between the above approaches is to be brought beyond the approaches themselves, in the form of a compilation of some self-consistent model with phenomenological signals of the relevant phase transitions.

This is a quite delicate matter, especially as regards the chiral transition, which requires hunting some effects caused by the lowering of the $\pi$-meson mass (down to zero) in the presence of an enormous background. As for the color confinement/deconfinement transition, some conclusions can be derived from the composition and multiplicities of the secondary hadrons. We now turn to the discussion of two conceivable scenarios of nuclear medium evolution.

## 5.2 The direct phase transition scenario QGP → H: pro and contra

In this scenario, the question-marked domain on the $\mu_B T$-plane (see Fig. 2 and also Fig. 9a) is practically absent, or in other words, the space–time evolution pattern is as in Fig. 9b. In accordance with this scenario, the chiral symmetry is restored and a short-term QGP is formed in the mid-rapidity region very quickly after two heavy nuclei of high energy collide; after cooling down fast to the temperature $T = T_c$, the chiral phase transition occurs that results immediately in hadronization of the nuclear matter. In the course of further expansion, this medium finally cools down to the kinetic freeze-out temperature $T_f$, which marks the end of the interaction process.

Being motivated by the bag model predictions (see above) regarding the chiral transition and hadronization temperature (under the assumption that it is the same one[21]), this scenario has been the subject of careful phenomenological and theoretical analysis directed at seeking arguments for its confirmation. It was soon noticed [40] that the experimental data on the fractional mid-(pseudo)rapidity yield of different hadron species coming from heavy-ion collisions at the SPS/CERN accelerator at $E_L \simeq 160$ MeV (were the experiment was most representative: data, although of rather poor accuracy, on the yield of about 20 different hadron species were available) could be described rather well and in an unexpectedly simple manner by the quite attractive, at first glance, hypothesis that already at the temperature $T_{ch} \simeq 170$ MeV, the nuclear substance turns into something like an ideal gas of hadrons,[22] which preserves its hadronic composition later on. This observation was called 'the early chemical freeze-out' (with $T_{ch}$ being its temperature) and was understood as a direct phenomenological confirmation that two phase transitions — the chiral symmetry breaking (restoration) and setting in (ceasing) of color confinement — occur at the same temperature ($T_c = T_d \simeq 180 – 200$ MeV). The additional parameters introduced are the baryonic chemical potentials of nucleons and hyperons, which fix baryon-to-antibaryon ratios within the fireball at $T = T_{ch}$. They are defined as follows: $\mu_{B^i} = \sum_{B^i} \mu_{u,d,s}$ ($\mu_u \simeq \mu_d$; below, the label $i$ is omitted when referring to nucleons), see Table 2. Thus, the multiplicity of $i$-th baryon is $N_{B^i} \sim (m_{B^i}/T)^{3/2} \exp\left[(\mu_{B^i} - m_{B^i})/T\right]$. Multiplicities of antibaryons and mesons can be found similarly (recalling that $\mu_q = -\mu_{\bar{q}}$), but a more explicit formula (of type (7), see below) is to be used for pions because $T_{ch}/m_\pi \simeq 1$.

In fact, however, a number of very obliging assumptions must be adopted for making this scenario inherently self-consistent. The first one, the most vulnerable in the very physical essence, is the quite radical and rather curious assumption [40] that at the chemical freeze-out temperature, the hadrons are essentially modified in a hot and dense medium ('dryout' down to 1/20 of their ordinary volume[23]), but preserve their mass and identity. The second assumption is that by definition of the chemical freeze-out, the proton-to-antiproton ratio (as well as the baryon-to-antibaryon one) remains fixed as the fireball cools down to vanishing interaction (kinetical freeze-out). At the top SPS/CERN accelerator energy, this evolution lasts over the temperature interval $\Delta T \simeq 60 – 70$ MeV and takes a very long time, $\Delta t \simeq 10 – 15$ fm [44, 53]; therefore, the antiproton-to-proton ratio is reasonably expected to fall by at least about one order of magnitude due to the quite large annihilation cross section, $\sigma_{\bar{p}p} \simeq 50$ mb, unless the antiprotons are somehow regenerated to approximately the same degree (see, e.g. [45]). And, finally, the third assumption is that chemical freeze-out follows the chiral phase transition almost immediately — otherwise the question (unessential for describing hadronic multiplicities themselves but, nevertheless, quite relevant to the understanding of evolution on the whole) remains open: what is the state of nuclear matter in between these events? In this connection, the chiral transition (and hence the hadronization) at the top SPS/CERN energy is usually expected to happen at $T_c \simeq 190$ MeV, although this is not quite in agreement with the lattice calculation results ($T_c \simeq T_d \simeq 154$ MeV, see above).[24] It is worth mentioning

---

[20] At the same time, exactly this equation prescribes the fireball convective (collective) transverse expansion rate and thus predetermines (side by side with $T_f$) the transverse distributions of final particles.

[21] The lattice calculations are invoked in favor of this version, although they predict a temperature at least 40–50 MeV lower than what is used in this scenario.

[22] These words imply that the relevant formulas are valid (with some small corrections accounting for the proper size of strongly in-medium modified hadrons) even in the presence of interaction.

[23] This is necessary for making the excluded volume small compared to the total one. Direct estimates — quite similar to what is made below in formula (7) for valons — show that weakly modified hadrons cannot form a nearly ideal gas because wave functions of the ordinary hadrons would essentially overlap at the relevant particle densities.

[24] This distinction can hardly be put down to an abnormally large mass of 'lattice quarks' since this factor brings a rather opposite effect (see [24]).

**Table 2.** Comparison between theory and experiment (the ratio of the particle yields $N_i/N_j$)

| $N_i/N_j$ | AGS Au+Au, $E_L = 11$ GeV/NN ($\sqrt{s_{NN}} \cong 5$ GeV) | | | SPS Pb+Pb, $E_L = 160$ GeV/NN ($\sqrt{s_{NN}} \cong 17.5$ GeV) | | | | RHIC Au+Au ($\sqrt{s_{NN}} = 200$ (130) GeV) | | | | LHC Pb+Pb ($\sqrt{s_{NN}} = 5.5$ TeV) | |
|---|---|---|---|---|---|---|---|---|---|---|---|---|---|
| | experiment | $M_1$ | $M_2$ | experiment | $M^*$ | $M_1$ | $M_2$ | experiment | $M^*$ | $M_1$ | $M_2$ | $M_1$ | $M_2$ |
| $p/\pi$ | 1 | 0.78 | 0.86 | 0.228 | 0.238 | 0.209 | 0.240 | 0.126 | 0.124 | 0.106 | 0.110 | 0.10 | 0.08 |
| $\bar{p}/p$ | $5 \times 10^{-4}$ | $4.7 \times 10^{-4}$ | $4.7 \times 10^{-4}$ | 0.067 | 0.055 | 0.084 | 0.080 | 0.632 | 0.629 | 0.628 | 0.628 | 1 | 1 |
| $K^+/\pi^+$ | 0.175 | 0.196 | 0.177 | 0.16 | | 0.165 | 0.17 | 0.172 | | 0.165 | 0.18 | 0.119 | 0.107 |
| $K^-/\pi^-$ | 0.034 | 0.044 | 0.035 | 0.085 | | 0.106 | 0.091 | 0.149 | 0.145 | 0.146 | 0.15 | 0.119 | 0.107 |
| $K_s^0/\pi^-$ | | 0.123 | 0.107 | 0.125 | 0.137 | 0.136 | 0.133 | | | 0.157 | 0.165 | 0.119 | 0.107 |
| $\eta/\pi^-$ | | 0.097 | 0.097 | 0.081 | 0.087 | 0.090 | 0.090 | | | 0.093 | 0.093 | 0.097 | 0.097 |
| $\Lambda/\pi^-$ | | 0.061 | 0.058 | 0.077 | 0.096 | 0.069 | 0.073 | 0.066 | 0.059 | 0.062 | 0.062 | 0.033 | 0.027 |
| $\Lambda/K_s^0$ | | 0.500 | 0.540 | 0.630 | 0.760 | 0.520 | 0.570 | | | 0.44 | 0.38 | 0.28 | 0.250 |
| $K^+/K^-$ | 5.14 | 4.450 | 4.890 | 1.85 | 1.90 | 1.54 | 1.89 | 1.156 | 1.118 | 1.125 | 1.211 | 1 | 1 |
| $\bar{\Lambda}/\Lambda$ | | 0.001 | 0.001 | 0.131 | 0.10 | 0.103 | 0.112 | 0.77 | 0.753 | 0.675 | 0.730 | 1 | 1 |
| $\Xi^-/\Lambda$ | | 0.090 | 0.093 | 0.101 | 0.110 | 0.109 | 0.107 | | 0.123 | 0.089 | 0.093 | 0.121 | 0.121 |
| $\Xi^+/\bar{\Lambda}$ | | 0.478 | 0.478 | 0.188 | 0.185 | 0.210 | 0.2 | | 0.145 | 0.099 | 0.101 | 0.121 | 0.121 |
| $\Xi^+/\Xi^-$ | | 0.002 | 0.002 | 0.232 | 0.228 | 0.2 | 0.2 | 0.82 | 0.894 | 0.797 | 0.798 | 1 | 1 |
| $\Omega^+/\Omega^-$ | | 0.013 | 0.013 | 0.383 | 0.53 | 0.385 | 0.382 | | 0.898 | 0.941 | 0.941 | 1 | 1 |
| $\phi/\pi^-$ | | 0.008 | 0.0076 | 0.021 | 0.019 | 0.013 | 0.017 | 0.025 | 0.02 | 0.024 | 0.024 | 0.018 | 0.018 |
| $\Delta^2$ | | 0.17 | 0.035 | | 0.36 | 0.42 | 0.4 | | 0.047 | 0.041 | 0.058 | — | — |
| $M^*$ | $\mu_B/T_{ch} \cong 4.3$ ($\mu_B \cong 540$ MeV) $\mu_S/T_{ch} \cong 0.59$ ($\mu_S \cong 75$ MeV) $T_{ch} \cong 125 \pm 6$ MeV | | | $\mu_B/T_{ch} \cong 1.58$ ($\mu_B \cong 266$ MeV) $\mu_S/T_{ch} \cong 0.42$ ($\mu_S \cong 71$ MeV) $T_{ch} \cong 168 \pm 2$ MeV | | | | $\mu_B/T_{ch} \cong 0.26$ ($\mu_B \cong 46$ MeV) $\mu_S/T_{ch} \cong 0.07$ ($\mu_S \cong 13$ MeV) $T_{ch} \cong 174 \pm 7$ MeV | | | | | |
| $T_f$ | $\cong 120$ MeV | | | $115 \pm 10$ MeV | | | | $105 \pm 10$ MeV | | | | | |
| $M_1, M_2$ | $\mu_B/T_d \cong 4.32$ $\mu_S/T_d \cong 0.59$ $T_d \cong 115 \pm 10$ MeV | | | $\mu_B/T_d \cong 1.47$ $\mu_S/T_d \cong 0.48$ $T_d \cong 115 \pm 10$ MeV | | | | $\mu_B/T_d \cong 0.28$ $\mu_S/T_d \cong 0.03$ $T_d \cong 115 \pm 10$ MeV | | | | $\mu_B/T_d \cong 0$ $\mu_S/T_d \cong 0$ $T_d \cong 115$ MeV | |

Note. $M^*$ refers to the model suggested and developed in Refs [40–42]. $M_1$ and $M_2$ are two versions of our model [50] with $\langle j_1 \rangle = 0.5$ and $0.7 \leqslant j_2 \leqslant 1$ (see the text). All the values of $\Delta^2$ were calculated by the authors in accordance with the cited experimental data (averages over the results obtained by different RHIC collaborations are used). $T_f$ is the kinetic freeze-out temperature calculated within the thermo-hydrodynamic model [41]. The reliable experimental observation that $N_{\pi^+} \simeq N_{\pi^-}$ was sometimes used for the ratios $N_i/N_j$ in the left column. Moreover, the identities $N_p/N_{\pi^-} \equiv (N_p/N_{\bar{p}})(N_p/N_{\pi^-})$, $N_{K^+}/N_{\pi^+} \equiv (N_{K^+}/N_{K^-})(N_{K^-}/N_{\pi^+})$, and $N_\phi/N_{\pi^-} \equiv (N_\phi/N_{h^-})(N_{h^-}/N_{\pi^-})$ for RHIC with actually measured right-hand-side ratios were used for obtaining the 'experimental' left-hand-side ones.

that the physical interpretation of the temperature $T_{ch}$ is somewhat obscured because, unlike all the other critical temperatures named above, it has nothing to do with phase transitions in the medium under consideration.

In a similar way, [41, 42, 46] the data of analogous — although by far less informative — experiments made at the SPS/CERN (at the lower energy, $E_L = 40$ GeV/NN), AGS/BNL, and SIS/GSI accelerators, as well as at the new RHIC/BNL accelerator were interpreted. [25] We note that the decrease in the chemical freeze-out temperature at lower energies ($T_{ch} \simeq 120$ and 60 MeV for AGS/BNL and SIS/GSI, respectively), which is seemingly favorable for this model, does not rehabilitate the second assumption above because of a sharp grow in baryonic chemical potential and an increase in the $\bar{p}p$-annihilation cross section. In addition, a problematic character of the third assumption (especially true for SIS/GSI) is highlighted: if it is adopted, then one has to agree that the chiral symmetry is restored at quite low temperatures even at $\mu_B < m_N$ (cf. the curves of the chiral phase transition predicted by the bag model [26] and of the early chemical freeze-out in Fig. 10). Although no strict objections have been put forward against such a pattern, a feeling of some dissatisfaction nevertheless remains [48].

### 5.3 The scenario with two phase transitions, QGP → Q$\pi$K → H: advantages and problems
This scenario manifestly implies the presence and significant role of the question-marked domain in the $\mu_B T$ plane (see

---

[25] To avoid unnecessary details, in Table 2 below, we present only the results related to the RHIC/BNL data [42].

[26] A similar behavior was also predicted by the lattice calculations [40, 47], however, they are still far from being quite reliable.

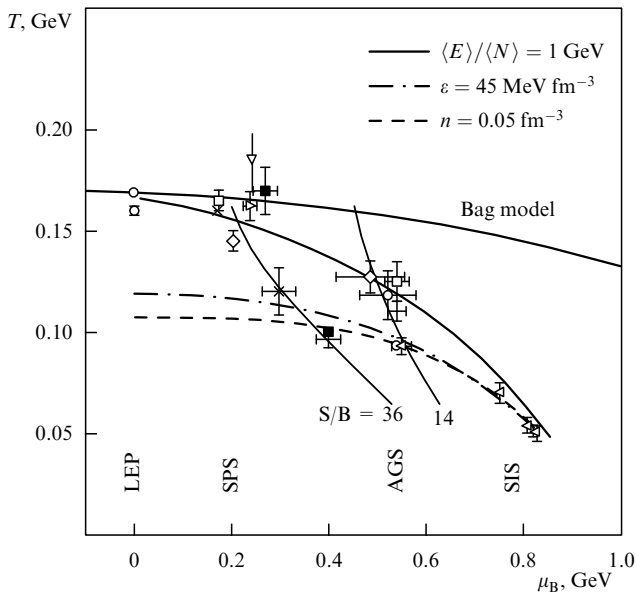

**Figure 10.** General appearance of the kinetic and early chemical freeze-out curves in the $\mu_B T$-plane. The 'experimental points' are obtained from the data of the direct measurements by their adaptation to the suggested model (see Ref. [46] and references therein). The solid curve interpolates the chemical freeze-out data (under the condition that the mean hadron energy $\langle E \rangle / \langle N \rangle = 1$ GeV); the dashed and dot-dashed curves interpolate the kinetic freeze-out data under the assumptions that it goes on either at a fixed particle density $n$ or at a fixed energy density $\varepsilon$, respectively. The curves crossing the above ones trace the cooling under the assumption that the specific entropy of baryon is conserved. In addition, the chiral phase transition curve is shown in accordance with the predictions of the bag model for the direct transition QGP $\rightarrow$ hadrons.

Fig. 2 and Fig. 9a), i.e., the space–time evolution pattern shown in Fig. 9c.

Certainly, in accordance with its physical motivation, the scenario under discussion shows a propensity for the version of the bag model allowing for an intermediate phase, but it has the advantage of being less restrictive and implying no appeal to an unknown EoS of the hadronic matter at high temperature and density. It is more tolerant to the admissible values of the deconfinement (hadronization) temperature $T_d$, which is predicted (see below) to be actually at least 30–40 MeV lower than in any version of the bag model. At the same time, the initial stages of fireball evolution are described in the conventional way: almost immediately after heavy nuclei of high energy collide, a short-lived QGP is formed in the mid-(pseudo)rapidity region, which quickly cools down to the temperature $T_c$ at which the chiral symmetry becomes broken ($T_c \simeq 200$ MeV for SPS/CERN at $E_L \simeq 160$ MeV/NN and for RHIC/BNL).

From here, the paths diverge. We suggest that the chiral phase transition does not result at once in the formation of the ordinary hadronic matter, but instead a specific and quite long-lived color-conducting Q$\pi$K phase occurs in which pions and kaons are the only hadron species that can survive. They stay in chemical and thermal equilibrium with the deconfined valons, which turn out to occupy a great share of the degrees of freedom because of large weight factors (see below). This primarily applies to the light valons Q(q) ($m_{Q(u,d)} \equiv m_{Q(q)} \simeq 330$ MeV) and strange valon Q(s) ($m_{Q(s)} \simeq 480$ MeV) — just those hypothetical particles that were widely and successfully employed in theoretical models, first of all, in the AQM before the QCD was elaborated.

As for all the other hadron species, they have too large masses and, therefore, are unstable within a color-conducting medium with respect to decay into valons, especially under collisions (see, e.g., Refs [49, 58] where the $\rho$-meson width broadening along with the increase in the nuclear matter density was analyzed in more detail).

While expanding and cooling down within the temperature interval from $T_c$ to $T_d$, the nuclear medium is steadily (and slowly) enriched with pions and kaons at the expense of Q$\bar{Q}$-annihilation. This (color-conducting) phase lasts until the total and quick confinement of all remaining valons — freezing of their degrees of freedom and total hadronization of the nuclear matter — sets in due to the lowering of the particle density. It is seemingly evident that this (second) phase transition occurs when the substance has become so rarefied that the color-screening length $L_{cs}$ (approximately equal to the mean spacing of valons) becomes nearly equal to the confinement radius. The analysis of the aforesaid experimental data on the fractional multiplicities of hadronic species shows (see the next section) that the relevant temperature [27] turns out to be quite close to that of the kinetic freeze-out, see Table 2 and Fig. 12.

An important peculiarity of this scenario is that no correlation exists between the hadronization process and the formation of QGP at the early stage of interaction: hadron production must go on in just the same manner if the initial interaction energy is still insufficient for excitation of the QGP phase but is high enough to form the equilibrium Q$\pi$K-phase (it might also be referred to as the valonic plasma). This prediction can be subjected to the direct experimental test in accordance with the above argument regarding the relatively small energy necessary for the valonic deconfinement.

Almost all the above disparities emerging from the assumption of the direct QGP–hadron transition seem to be resolved within the framework of this scenario. First of all, an unambiguous answer is given to the question of what state the matter is in between the chirally symmetric and hadronic phases, instead of a rather conflicting hypothesis on early chemical freeze-out (within this scenario, there is no room for this notion at all). The valons produced just after chiral symmetry breakup are quite probably best considered as a nearly ideal gas because, unlike hadrons, they are actually particles of a small size [10], $r_Q \simeq 0.3$ fm. Indeed, their Boltzmann ideal-gas density can be estimated at $T \simeq 170$–200 MeV as

$$\frac{GT^3}{\pi^2}\left[2\left(\frac{m_{Q(q)}}{T}\right)^2 K_2\left(\frac{m_{Q(q)}}{T}\right)\cosh\left(\frac{\mu_{Q(q)}}{T}\right) + \left(\frac{m_{Q(s)}}{T}\right)^2 K_2\left(\frac{m_{Q(s)}}{T}\right)\cosh\left(\frac{\mu_{Q(s)}}{T}\right)\right] \simeq 0.5 \text{ fm}^{-3}, \quad (7)$$

where $G = 2N_c(2S_q + 1)$, $N_c = 3$ is the number of flavors, and $S_q = 1/2$ is the quark (valon) spin. This estimate involves the chemical potential values $\mu_{Q(q)}$ and $\mu_{Q(s)}$ for the Q(q) and Q(s) valons that are typical of heavy-ion collisions. Thus, the proper (excluded) volume of valons — $(4\pi/3)(0.3 \text{ fm})^3 \simeq 0.1 \text{ fm}^3$ per valon — would take only about 5% of the total one even at such high temperatures. This quite optimistic estimate evidently worsens if a necessary chemical equilibrium admixture of 'big' particles — pions and kaons — is taken into account. Fortunately, this hadronic fraction is proven [50] to not exceed 25% and, therefore, this three-component Q$\pi$K-state can still be considered a gas, [28] although it becomes somewhat more questionable whether this gas is of a quasi-ideal nature. Furthermore, because the temperatures $T_d$ and $T_f$ follow very shortly one after the other, free scattering starts very quickly after hadronization and, thus, the problem of the baryon–antibaryon annihilation becomes less acute, although it must be taken into account in considering the AGS/BNL data (but no corrections of the order of magnitude are then expected, see below). That is why this scenario is somewhat more economical than the above version of the bag model predicting a noticeably higher value of the temperature $T_d$, which can hardly be associated unambiguously with the experimental data on the yields of different hadron species. [29] We note the transparent physical interpretation of the fact that the temperature interval $T_d - T_f$ must be very small: the hadrons are formed as discernible and stable physical objects only very shortly before they become free, which clearly correlates with the short-range nature of the nuclear forces.

---

[27] Certainly, if a conventional phase transition, but not a crossover blurring the temperature, takes place.

At the same time, one vulnerable point of this tempting scenario should be emphasized: the notion of the valon itself as a real quasi-particle has not been embedded rigorously into the body of QCD [51] yet, which is not very surprising because it may only be a physical entity of an especially non-perturbative nature.

**Fractional yields of different hadron species: a general pattern of calculation.** One can see from the above discussion of hadron production that this question is inseparably interspersed with the substance of both scenarios. The currently prevailing scenario was mentioned to be quite straightforward — this is its evident merit — and the corresponding formal algorithm of calculation is quite transparent. At the same time, the scenario allowing for an intermediate phase requires concrete definition, and we now present a schematic example of how this approach 'works', being applied to the description of the experimental data on production of various hadronic species in the existing accelerator experiments (see Ref. [50] for more details). We have already noted that although the three-component valon–pion–kaon medium formed as a result of chiral symmetry breaking is certainly gaseous, it is not *a priori* guaranteed that this gas is close to the ideal gas in its properties. That is why we rely below, as far as possible, on a more general consideration based exclusively on the chemical kinetics (the detailed balance equation) and then contrast the results against those obtained under the assumption that this gas is ideal near the hadronization temperature. It follows from Table 2 that the results obtained by these two methods are quite compatible.

In the course of the phase transition Q$\pi$K $\to$ H at $T = T_d$, all the hadrons are produced, generally speaking, via the same mechanism — annihilation of the corresponding valons (see Fig. 11). Because the correlation length becomes large compared to the mean particle spacing (it is even 'infinite' if the second-order phase transition occurs), the many-particle interaction dominates at this stage of evolution. It is also reasonable to assume that all the produced hadrons survive in such conditions because their disintegration is suppressed by the confinement mechanism.

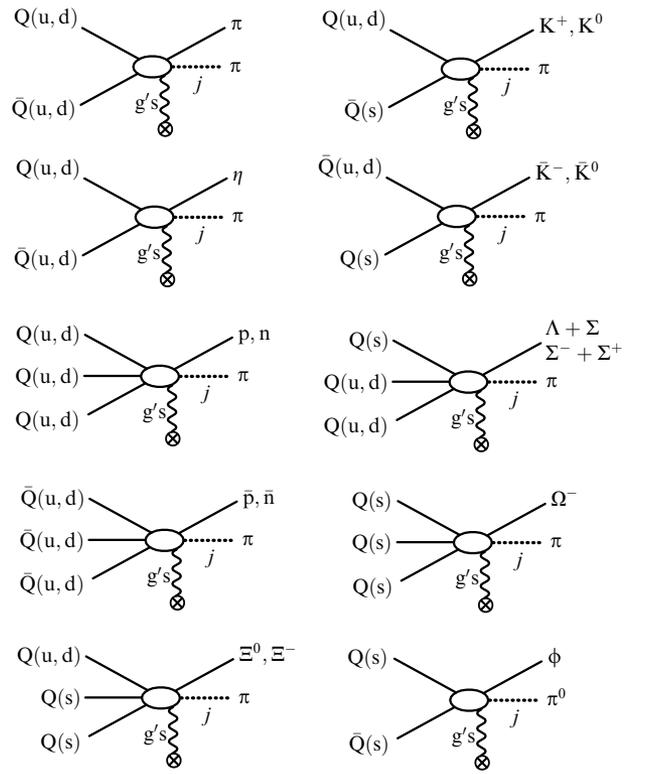

**Figure 11.** A scheme of color annihilation forming hadrons at $(T \simeq T_d)$. The dotted and wavy lines refer to pions and gluons, respectively, which accompany the 'principal' hadron creation (gluons are absorbed by the medium).

Certainly, in calculating the yield of pions and kaons, one has to allow for the ones, $n_\pi$ and $n_K$, that were accumulated in the Q$\pi$K-phase before the moment of global hadronization. Their number is determined by the detailed balance equation in the reactions

$$Q(u,d) + \bar{Q}(u,d) \leftrightarrow \pi + \pi,$$
$$Q(u,d) + \bar{Q}(s) \leftrightarrow K^+(K^0) + \pi,$$
$$Q(s) + \bar{Q}(u,d) \leftrightarrow K^-(\bar{K}^0) + \pi,$$
$$Q(u,d) + K^-(\bar{K}^0) \leftrightarrow Q(s) + \pi,$$
$$\bar{Q}(u,d) + K^+(K^0) \leftrightarrow \bar{Q}(s) + \pi.$$

In addition, some amount of pions still has to be added because, owing their small mass, they can accompany the creation of other hadrons, see Fig. 11.

The number of pions produced in the course of hadronization via the same mechanism as all the other hadron species can be estimated by tracing the fate of a valon $Q(q)$ (antivalon $\bar{Q}(q)$) under hadronization of the nuclear matter. During the mean free time, it is annihilated in a collision with an antivalon (valon) with the probability $n_{\bar{Q}(q)}/n$ ($n_{Q(q)}/n$), with $n_{Q(q)}$ ($n_{\bar{Q}(q)}$) being the number of $Q(q)$-valons ($\bar{Q}(q)$-antivalons), and therefore the number of such collisions equals $n_{Q(q)} n_{\bar{Q}(q)}/n$. Each collision results in the final state $\pi + X$, where '$X$' may also include a number $j$ of additional pions (the same is true for the processes of other hadron production), see Fig. 11. Thus, the total number of negative pions (nearly 1/3 of the total number of pions) is equal to (also accounting for the 'primary' pions and kaons)

$$N_{\pi^-} \simeq \frac{1}{3}\left[n_\pi + \frac{(1+\langle j \rangle)\, n_{Q(q)} n_{\bar{Q}(q)}}{n}\right]$$
$$+ \frac{\langle j \rangle}{3}\left[N_B + N_{\bar{B}} + (N_K - n_K)\right], \qquad (8)$$

where $N_B$ ($N_{\bar{B}}$) and $N_K$ is the total yield of baryons (antibaryons) and K-mesons and $\langle j \rangle$ is the mean value of $j$, which is easily estimated to be $0 \leqslant j \leqslant 1$ because of the limited mean phase space volume (that is why the results are weakly dependent on the value of $\langle j \rangle$ within this domain).

---

[28] This is expected to be, all the more, the case near the relatively low hadronization temperature $T_d$, which is especially significant in the context of the problems under discussion

[29] By the way, this point may be useful for reducing the arbitrariness in the choice of the bag model parameters (see above).

The yields of other mesons can be easily calculated by similar combinatorial considerations and/or estimations of the corresponding cross sections; the baryonic yields can be calculated similarly. For example,

$$N_\phi \simeq \frac{n_{Q(s)} n_{\bar{Q}(s)}}{n}, \qquad N_p \simeq N_n \simeq \frac{n_{Q(q)}^3}{2n^2(1 + 2n_{Q(q)}^2/n^2)}, \qquad (9)$$

and so on. An exception is given by the ratio $N_{\bar{p}}/N_p$ in the AGS/BNL experiments, because the high proton density $\mu_B/T_d = 3\mu_{Q(q)}/T_d \simeq 4.2$ [30] means that one can no longer neglect the antiproton annihilation in the course of hadronization. This follows directly from the fact that the experimental value [52] of the ratio $N_{\bar{p}}/N_\pi$ decreases along with the impact parameter and reaches its minimum for the central interactions. A simple (and, of course, crude) estimate of the possible effect is evident [50]; it can explain suppression of the antiproton yield approximately by a factor of two due to the $\bar{p}p$-annihilation in the narrow temperature interval $T_d \geqslant T \geqslant T_f$. This correction was subsequently taken into account.

The optimum values of the ratios $n_{Q(q)}/n_{\bar{Q}(q)}$, $n_{Q(s)}/n_{\bar{Q}(s)}$, and $n_{Q(q)}/n_{Q(s)}$ were determined by minimization of the sum of the squared deviations,

$$\Delta^2 \simeq \min\left[\sum_{i=0}^{i=k}\left(1 - \frac{a_{\text{th}}^i}{a_{\text{exp}}^i}\right)^2\right],$$

where $k$ is the number of different species yields considered. [31]

To use the above optimization results for obtaining an estimate of the corresponding hadronization temperature, one still has to assume that $Q\pi K$-gas is nearly ideal in the vicinity of this temperature. One then obtains

$$\frac{n_{Q(q)}}{n_{\bar{Q}(q)}} = \exp\left(\frac{2\mu_{Q(q)}}{T}\right), \quad \frac{n_{Q(s)}}{n_{\bar{Q}(s)}} = \exp\left(\frac{2\mu_{Q(s)}}{T}\right)$$

and

$$\frac{n_{Q(q)}}{n_{Q(s)}} \simeq \left(\frac{m_{Q(q)}}{m_{Q(s)}}\right)^{3/2} \exp\left[\frac{m_{Q(s)} - m_{Q(q)}}{T}\right]$$
$$\times \exp\left[\frac{\mu_{Q(q)} - \mu_{Q(s)}}{T}\right]. \qquad (10)$$

---

[30] We note incidentally that additivity of the valonic chemical potentials within the hadron seems to be by far more understandable than the same assumption attributed to the current quarks (see above).
[31] In Refs [40–42], the conventional procedure of $\chi^2$ minimization was also performed, which is, strictly speaking, more justified. However, the current data precision is still too low to distinguish between the two methods within the experimental errors. That is why we here present the above procedure only, although it is not quite correct.

Having assumed this, we obtain nearly the same hadronization temperature, $T_d \simeq 115 \pm 10$ MeV, for all of the experiments — namely, for AGS/BNL, SPS/CERN, and RHIC/BNL. The pion fraction in the $Q\pi K$-phase is also confirmed to be rather small, as was already mentioned, even just before the hadronization: 0.13 for AGS/BNL and 0.22 for SPS/CERN, RHIC/BNL, and LHC/CERN (the kaon fraction is still far lower, but it contributes noticeably to the total multiplicity observed).

But the scarcity and rather low quality of the experimental data — first and foremost, the AGS/BNL data — prevents us from insisting that the temperature $T_d$ is independent of the chemical potential with some quantitative reliability. Actually, the AGS/BNL hadronization temperature may be somewhat lower (or not even attained at all) in the relevant experiments. [32] An important point, however, is that this scenario predicts only the hadronization temperature, which is well below the chiral transition temperature (180–200 MeV) attributed to the in-fireball nuclear matter at RHIC/BNL and the top SPS/CERN energies.

The obtained results are collected in Table 2 (see column $M_1$) as well as in Fig. 12, where they are compared to the predictions [46] of the early chemical freeze-out model (see Fig. 10). We note again that similar results are obtained if $Q\pi K$-gas is assumed to be ideal from the very beginning (see column $M_2$ in Table 2) instead of using the detailed balance equation for its description. The predictions for future experiments at LHC/CERN were obtained for the same hadronization temperature $T_d$ and zero values of all chemical potentials: $\mu_{Q(i)} = 0$ (and, certainly, $\mu_B = 0$, see Figs 2 and 12).

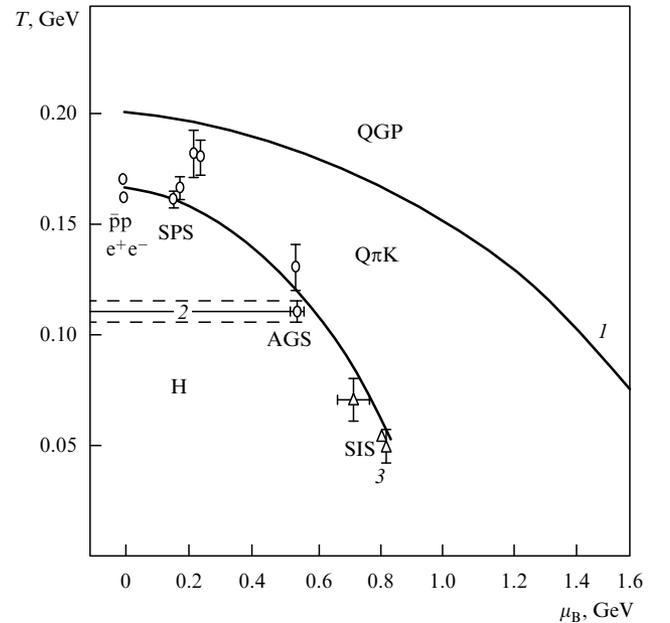

**Figure 12.** Comparison of two hadronization scenarios — with and without the intermediate phase $Q\pi K$. Curve *1* refers (rather qualitatively, see Ref. [3]) to the chiral phase transition QGP → $Q\pi K$ in the scenario with the intermediate phase incorporated; strip *2* between two horizontal dashed lines shows the temperature interval around $T_d = 115$ MeV predicted for the hadronization process $Q\pi K$ → H within the same scenario [50]. The early chemical freeze-out curve *3* is borrowed from Fig. 10.

---

[32] The recent analysis [53] of the SPS/CERN experimental data at energies $E_L \geqslant 20$ GeV/NN ($\sqrt{s_{NN}} \geqslant 6.5$ GeV) probably showed that some qualitative changes in the features of interaction necessarily attributed to a phase transition are only observed at $\sqrt{s_{NN}} \geqslant 7$ GeV and hence certain doubts arise regarding whether the AGS/BNL energy ($\sqrt{s_{NN}} \simeq 5$ GeV) was sufficiently high for color deconfinement, i.e., for 'jumping' over curve *1* in Fig. 2.

### 5.4 Dilepton ($e^+e^-$-pairs) production

Another important process that has long been scrutinized experimentally and is sufficiently well described theoretically within each of the two scenarios is the direct $e^+e^-$-pair production in collisions of relativistic nuclei. Long before the QCD was elaborated and, what is more, before the corresponding large-scale experiments were put on the agenda, one of us (E L F) called attention [54] to the possibility of direct photon and dilepton generation in the course of such interactions and noticed that the relevant observations may bring to light the essence and properties of nuclear matter under extreme conditions. In this connection, two mechanisms are to be distinguished: electromagnetic processes at very high temperature (in the QGP-phase) — these are mainly (current) quark + gluon → (current) quark + (real or virtual) photon (and the corresponding cross reaction $q\bar{q} \to g\gamma$) — and electromagnetic processes after the chiral phase transition involving hadrons and valons (if the intermediate phase exists). The former ones should affect the spectra of photons with high transverse momenta and dileptons with large invariant masses (enriching both spectra compared to what would be expected if no QGP were formed at all), and thus may emerge as the most direct signal of QGP formation; the latter ones are responsible for production of rather soft photons and low-mass dileptons and provide information on the duration of fireball evolution from the chiral phase transition until the kinetic freeze-out. As a matter of principle, either of them is to be described within the framework of the thermodynamic approach [2, 55] which, however, being quite good for describing the main qualitative trends, is still insufficient [33] for predicting quantitative results that would be unambiguously compatible with the experimental data. Some very nontrivial peculiarities of the $e^+e^-$-pair spectrum with low invariant masses [56, 57] are described, on the whole, quite successfully within both the direct phase transition scenario QGP → H [58] and the scenario involving the intermediate phase, QGP → $Q\pi K$ → H [60]. Nevertheless, a certain noticeable distinction is also seen in the relevant estimates.

It is common knowledge that the attempt to link the data on $e^+e^-$-pair production in $pA$- and $AA$-interactions by multiplying the former ones by the factor $A$ results in a direct contradiction with the experimental data within the invariant mass interval 250 MeV < $M_{ee}$ < 700 MeV [56, 57], see Fig. 13: the dilepton yield turns out to be much greater than expected (the maximum discrepancy fell on the mass $M_{ee} \simeq 500$ MeV, where the enhancement factor is about 5). Most probably, this way of counting also results in a considerable theoretical underestimation of dilepton yield at the masses 1 GeV $\leqslant M_{ee} \leqslant$ 1.5 GeV (some discrepancy already seems noticeable, although no definite statement can be made because of a very poor experimental accuracy). The above loud disparity is broadly explained as being caused by the quite long hadronic phase through which the nuclear matter evolved (see above) when multiple $\pi$-meson collisions produced the well-modified (broadened) in-medium vector-meson resonances (predominantly $\rho$-mesons) which, in turn, decaying into the electron–positron channel, produced the enhanced yield of dileptons with the modified spectrum [49, 58, 59]. If one takes a proper medium density dependence of resonance broadening (in other words, if the desirable EoS is chosen) and assumes that the vector meson decay branching ratio into the dilepton channel remains unchanged, then, indeed, one can obtain a rather good fit of the measured dileptonic spectrum [58] (see thin solid curves in Figs 13 and 14) at 250 MeV < $M_{ee}$ < 700 MeV, with the exception of the near-threshold region $M_{ee} \simeq 2m_\pi$ (see Fig. 14a).

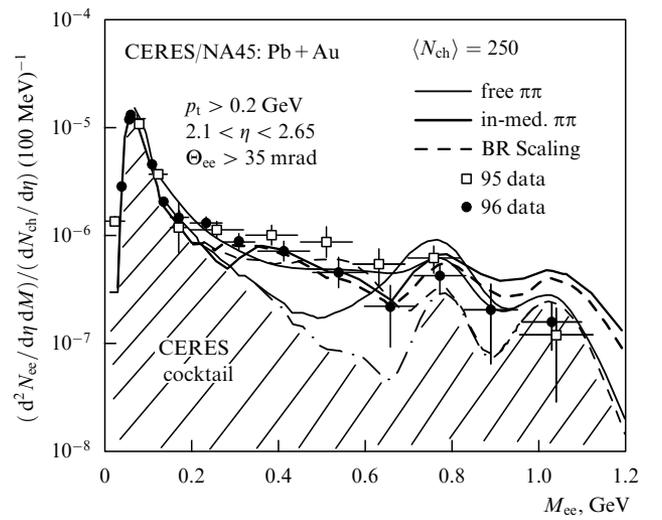

**Figure 13.** Comparing the models with the experimental data [56, 57] on the yield of dileptons ($e^+e^-$ pairs) with small invariant masses. The results obtained within the intermediate-phase scenario [60] (the solid and dashed bold curves refer to $m_\pi = 140$ MeV and $\tau/\langle t \rangle = 20$ and to the averaged in-medium pion mass $m_\pi = 100$ MeV, and $\tau/\langle t \rangle = 30$, respectively) are compared to the results obtained ignoring this phase [58] (the thin curves).

If we consider the long $Q\pi K$-phase instead of the long-lived hadron phase (just as we did in the above description of the secondary hadron fractional yields), then we obtain a theoretical explanation [60] of the dileptonic spectrum that is, at least, no worse, while seeming sometimes even somewhat better: it follows more accurately the experimental data at $M_{ee} \simeq 2m_\pi$ and $M_{ee} \geqslant 1$ GeV (see the bold curves in Figs 13 and 14a). Within this approach, dileptonic excess over the CERES-cocktail [34] results from their generation in the course of the enormous reiteration of the prompt $\pi\pi$- and $Q\bar{Q}$-collisions over the long evolution of the $Q\pi K$-phase. The combinatorial method of calculation (see Ref. [60] for this and other details) essentially coincides with what was done in calculating the yield of the different hadron species, and we therefore give only the final formula, in which a certain effective temperature (close to its averages in between the chiral transition and hadronization temperatures), $\langle T \rangle = 160$ MeV, and the corresponding mean free time,

---

[33] Only because of the lack of reliable methods for handling many-particle systems with strong interparticle interaction.

[34] This name is given to the domain below the dash-dotted curves in Figs 13 and 14 obtained by straightforward multiplication of the data on dileptonic yields in proton–nucleus interactions by $A$, when no nuclear fireball is produced at all.

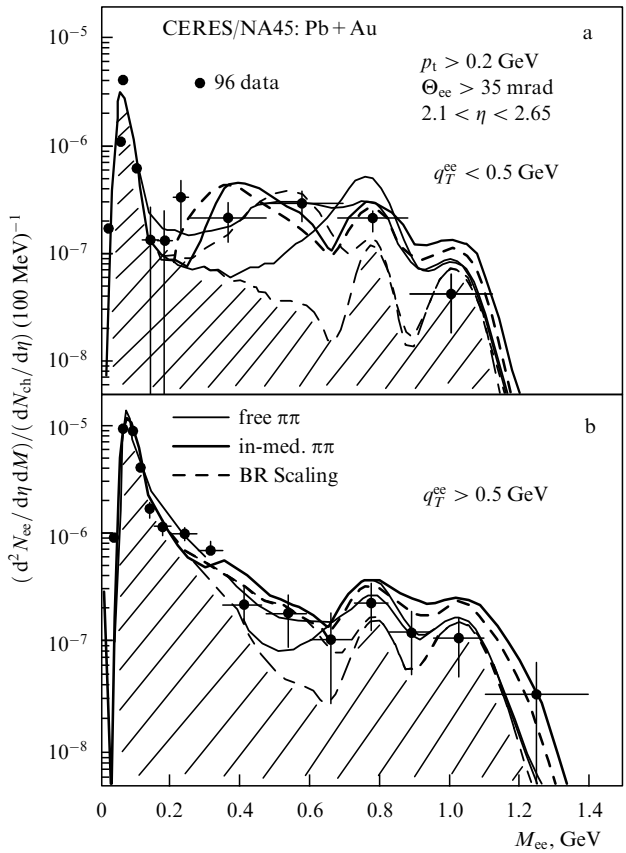

**Figure 14.** As in Fig. 13, but for two selected groups of the data ($q_T^{ee}$ being the dilepton ($e^+e^-$-pair) transverse momentum), see Refs [57, 60, 62].

$\langle t \rangle \simeq 0.5$ fm, are used instead of the current ones,

$$\frac{1}{N_{ch}} \frac{dN_{ee}}{dM_{ee}} \simeq \frac{0.1\lambda^2(1 + 0.6\lambda + \lambda^2)^{-1}}{4.7 + \lambda + 0.33\lambda^2} \frac{\tau}{\langle t \rangle}$$
$$\times \left[ \frac{dW_{\pi\pi}}{dM_{ee}} \frac{\sigma_{\pi\pi \to ee}}{\sigma_{\pi\pi}^{tot}} + 4 \frac{dW_{Q\bar{Q}}}{dM_{ee}} \frac{\sigma_{QQ \to ee}}{\sigma_{Q\bar{Q}}^{tot}} \right], \quad (11)$$

where $N_{ch}$ is the total number of charged hadrons (predominantly pions), $\lambda^2 = N_{Q(q)}/N_{\bar{Q}(q)} \simeq \exp(2\mu_{Q(q)}/T)$, $\tau$ is the Q$\pi$K-phase duration, $dW_i/dM_{ee}$ is the probability of finding the total 4-momentum squared $(p_1 + p_2)^2 = M^2$ under a random collision of two pions or valons, $\sigma_{Q(q)\bar{Q}(q)}$ is half the sum of the cross sections for Q(u)$\bar{Q}$(u) and Q(d)$\bar{Q}$(d) annihilation into $e^+e^-$, and $\sigma_{\pi\pi}^{tot}$ and $\sigma_{Q\bar{Q}}^{tot}$ stand for the corresponding total cross sections.[35]

Two qualitative results follow immediately from Eqn (11). First, the only $N_{ch}$-dependent factor in its right-hand side is the ratio $\tau/\langle t \rangle$, where the functional dependence of the time $\tau$ is varied from $\tau \sim V \sim N_{ch}$ for the one-dimensional (longitudinal) expansion to $\tau \sim V^{1/3} \sim N_{ch}^{1/3}$ for the three-dimensional (spherical) one ($V$ is the fireball volume just before hadronization). Thus, on the one hand, Eqn (11) allows the correlation observed in the SPS/CERN experiments, $N_{ee} \sim N_{ch}\tau \sim N_{ch}^2$, which implies a quasi-one-dimensional expansion, while, on the other hand, it predicts a weakening of the dependence of $N_{ee}$ on $N_{ch}$ as the collision energy grows (RHIC/BNL, LHC/CERN) because the transverse expansion (sphericity) is then expected to become more pronounced. Second, the (relative) dileptonic excess is expected to fall at considerably lower energies (insufficient for making the QGP) due to the increase in the value of $\lambda$ and the decrease in the ratio $\tau/\langle t \rangle$.

At the same time, an obvious drawback of this method should be mentioned: the total valon–antivalon cross section enters Eqn (2), which, of course, cannot be measured directly (something like a payment for eliminating the necessity of considering the equally obscure properties of vector mesons within hot and/or dense media). Its value and energy dependence were estimated under the standard assumptions that allow using the data on the p$\bar{p}$-interaction cross section at the appropriately low energies. The more accessible but still unmeasured cross section of the low-energy $\pi\pi$ interaction was chosen similarly using the analogy with the $\pi$p scattering (see Ref. [60] for more details).

## 6. Concluding remarks

It is probably most reasonable to paraphrase the aforesaid as follows. Below the chiral symmetry breaking temperature $T_c$, nuclear matter can be described in two ways: it can be considered either as a very hot and/or dense 'hadronic liquid' or as a certain specific state useless for hadronic survival (with the exception of pions and kaons) and dominated by the valon (not the hadron) degrees of freedom,[36] the hadrons coming to life only after cooling down to the hadronization temperature $T_d$. Certainly, one can choose either of these approaches. However, the former one implies, at least, the knowledge of the relevant nontrivial EoS that, in fact, is to be postulated. As regards the latter approach, which we are attempting to suggest, its most important point is that the valon–pion–kaon substance produced at the stage of chiral symmetry breaking can, most probably, be actually considered as a gas from the very beginning (in contrast to hadrons), because valons are in fact quasi-particles of a small size, whereas the pionic and kaonic fractions are insignificant. This gas might be referred to, in a certain sense, as valonic plasma, because it allows free propagation of color particles.

*A priori*, we have no decisive arguments in favor of either of the two versions. The direct lattice calculations are undoubtedly promising and will possibly provide all the answers in the future, but the currently available results are still limited to $\mu_B = 0$ and, furthermore, it is rather difficult to weigh their sensibility and to bring them nearer to reality (in particular, to handle current quarks with realistic masses) because the computers are not fast enough by far. Nevertheless, one can even now notice an obvious trend in the lattice deconfinement temperature becoming lower along with allowing for the s quarks and making the 'lattice' u and d quarks more realistic. Thus, it is conceivable that a noticeable distinction in the estimates of this temperature given by the lattice calculations, on the one hand, and within the indirect phase transition scenario, QGP $\to$ Q$\pi$K $\to$ H, on the other hand, will be reduced in time due to the further lowering of the lattice temperature.

---

[35] The strange valons drop out from the right-hand side of Eqn (11) because their contribution does not exceed 5–6% [60], which is within both the experimental error and theoretical accuracy.

[36] The presence of massive valons clearly correlates with the breaking of the chiral symmetry.

For now, however, the only hope of making a reasonable choice relies on comparing the possibility for noncontradictive descriptions of the experimental data within the frameworks of the two approaches. In this connection, the observations of primary interest are those that make it possible to judge somehow the properties of nuclear matter as it evolves from the chiral symmetry breaking to the cessation of interaction between hadrons (kinetic freeze-out) and beginning of their free scattering. That is just why, among the numerous and manifold theoretical problems and sometimes controversial experimental results, we have selected for discussion in more detail only those which are directly related to this question: possible color deconfinement at a rather low compression and/or temperature (and, thus, valonic deconfinement as well as the related inverse process — hadronization) and direct production of dileptons with low invariant masses.

In our opinion, the same reasoning calls primary attention to the purposeful experiments at relatively low energies, when the chirally symmetric phase is not yet attained and the quark–antiquark condensate survives. In this connection, the above-mentioned results [61] obtained at SPS/CERN in the course of energy scanning of $E_L$ from 20 to 40 GeV/NN are of particular interest. Searching for and discovering some effects indicative of short-term color deconfinement in nuclear interactions under such conditions would answer the question about the validity of the notion of constituent quarks (valons) as real quasi-particles, which, in turn, would be of essential importance in understanding the QCD structure on the whole.

**Acknowledgments.** The authors are indebted to M I Polikarpov for fruitful discussions of some questions related to lattice calculations.

This work was supported in part by the RF President grant for "Leading scientific schools" (no. NS.1936.2003.2) and by the Russian Foundation for Basic Research (project no. 03-02-16134).